\documentclass{aa}  

\usepackage{graphicx}
\usepackage{txfonts}
\usepackage{mathrsfs}
\usepackage[colorlinks,allcolors=blue]{hyperref}
\usepackage{breakurl}
\usepackage{subfigure}
\usepackage{stackengine}

\def \src {\mbox{J17407}}
\def \nustar {\mbox{\emph{NuSTAR}}}
\def \xmm {\mbox{\emph{XMM-Newton}}}
\newcommand{\chandra}{\textit{Chandra}}

\newcommand{\rosat}{ROSAT}

\newcommand{\swift}{\textit{Swift}}
\newcommand{\integral}{INTEGRAL}

\begin{document}

\title{X-ray flashes from the low-mass X-ray binary IGR~J17407$-$2808}
\titlerunning{X-ray flashes from the LMXB IGR~J17407$-$2808}

\author{L. Ducci
          \inst{1,2,4}
          \and
          C. Malacaria
          \inst{3}
          \and
          P. Romano
          \inst{4}
          \and
          E. Bozzo
          \inst{2}
          \and
          M. Berton
          \inst{5}
          \and
          A. Santangelo
          \inst{1}
          \and
          E. Congiu
          \inst{5}
          }

   \institute{Institut f\"ur Astronomie und Astrophysik, Kepler Center for Astro and Particle Physics, University of Tuebingen, 
              Sand 1, 72076 T\"ubingen, Germany\\
              \email{ducci@astro.uni-tuebingen.de}
              \and      
              ISDC Data Center for Astrophysics, Universit\'e de Gen\`eve, 16 chemin d'\'Ecogia, 1290 Versoix, Switzerland
              \and
              International Space Science Institute (ISSI), Hallerstrasse 6, 3012 Bern, Switzerland
              \and
              INAF -- Osservatorio Astronomico di Brera, via Bianchi 46, 23807 Merate (LC), Italy
              \and
              European Southern Observatory (ESO), Alonso de Córdova 3107, Casilla 19, Santiago 19001, Chile
            }

   \date{Received 1 March 2023; accepted 15 April 2023}

 
  \abstract
      {IGR~J17407$-$2808 is an enigmatic and poorly studied X-ray binary that was recently observed quasi-simultaneously with \nustar\ and \xmm. In this paper we report the results of this observational campaign. During the first 60~ks of observation, the source was caught in a relatively low emission state, characterised by a modest variability and an average flux of $\sim 8.3\times 10^{-13}$~erg~cm$^{-2}$~s$^{-1}$ (4$-$60~keV). 
      Afterwards, IGR~J17407$-$2808 entered a significantly more active emission state that persisted for the remaining $\sim$40~ks of the \nustar\ observation. 
        During this state, IGR~J17407$-$2808 displayed several fast X-ray flares, featuring durations of $\sim 1-100$~s and profiles with either single or multiple peaks. 
        The source flux in the flaring state reached values as high as $\sim 2\times 10^{-9}$~erg~cm$^{-2}$~s$^{-1}$ (4$-$60~keV), leading to a measured dynamic range during the \nustar\ and \xmm\ campaign of $\gtrsim10^3$.   
        We also analysed available archival photometric near-infrared data of IGR~J17407$-$2808 to improve the constraints available so far on the the nature of the donor star hosted in this system. Our analysis shows that the donor star can be either a rare K or M-type sub-subgiant or an K type main sequence star, or sub-giant star. Our findings support the classification of IGR~J17407$-$2808 as a low-mass X-ray binary.
        We discuss the source X-ray behaviour as recorded by \nustar\ and \xmm\ in view of this revised classification. 
        } 


   \maketitle
%

\begin{table*}[ht!]
\begin{center}
\caption{Summary of the X-ray observations.}
\vspace{-0.3cm}
\label{table log}
\resizebox{\columnwidth+\columnwidth}{!}{
\begin{tabular}{lcccccc}
\hline
\hline
\noalign{\smallskip}
Satellite &  observation ID  &        \multicolumn{2}{c}{Start time} &    \multicolumn{2}{c}{End time}     &               Net exposure                 \\
\hline
\noalign{\smallskip}
          &                  &          (UTC)       &   (MJD)        &          (UTC)       &     (MJD)    &                  (ks)                      \\
\noalign{\smallskip}
\hline
\noalign{\smallskip}
\nustar   &   30801013002    & 2022-09-13 15:06:09  &  59835.629     &  2022-09-14 23:21:09 &  	59836.973  &           FPMA: 55.93; FPMB: 55.45         \\
\noalign{\smallskip}
\xmm      &   0913200201     & 2022-09-13 14:29:08  &  59835.603     &  2022-09-14 03:18:08 &   59836.137  & \emph{pn}: 25.66; MOS1: 38.20; MOS2: 37.93 \\ 
\noalign{\smallskip}
\hline
\end{tabular}
}
\end{center}
\end{table*}

   \section{Introduction}

   IGR~J17407$-$2808 (hereafter: \src) is an X-ray binary discovered by the International Gamma-Ray Astronomy Laboratory
   (\integral) in 2004 during an enhanced emission activity that
   ended with a bright and fast flare achieving a peak flux of $\sim 9.5\times 10^{-9}$~erg~cm$^{-2}$~s$^{-1}$
   (20$-$60~keV) and featuring a duration of about one minute \citep{Kretschmar04, Gotz04, Sguera06}.
   \src\ has been observed so far by several past and currently operating X-ray facilities, including \rosat, \integral, \chandra, \swift,
   and \xmm\ \citep{Sidoli01, Kretschmar04, Ducci10, Mereminskiy20, Tomsick08, Heinke09, Romano11, Romano16}. 
  \src\ is frequently observed in a low emission state, characterised  by a 2$-$10~keV flux that varies from $\sim 1.7\times 10^{-13}$~erg~cm$^{-2}$~s$^{-1}$ to $\sim 2\times 10^{-12}$~erg~cm$^{-2}$~s$^{-1}$.
   In almost all pointed X-ray observations performed during the low emission state (endowed with typical exposure times of $\sim$10$-$20~ks), at least one short (50-400~s) and weak ($F_{\rm x}$[2-10~keV]$\approx 1-8\times 10^{-12}$~erg~cm$^{-2}$~s$^{-1}$) flare has been detected. 
   Brighter flares are rare, but in some cases peak fluxes as high as  $F_{\rm x}$[20-60~keV]$\approx 10^{-9}-10^{-8}$~erg~cm$^{-2}$~s$^{-1}$ have been measured. 
   The ``activity duty cycle'' (fraction of time spent by the source above $\sim 2\times 10^{-10}$~erg~cm$^{-2}$~s$^{-1}$ in 20$-$40~keV)
   is thus estimated at $\approx 0.05$\% \citep{Ducci10, Romano14}.
   In the soft X-ray domain (0.3$-$10~keV), the X-ray spectrum of \src\ is usually well fit with a simple absorbed power law model. 
   A hint for a curvature in the hard part of the source spectrum was reported by \citet{Romano16}, using simultaneous \swift/X-ray Telescope (XRT)
   and Burst Alert Telescope (BAT) data ($\sim 0.3-70$~keV) collected during a bright flaring event 
   (the measured unabsorbed flux was of $F_{\rm x}$[0.5-100~keV]$\approx 3.6\times 10^{-9}$~erg~cm$^{-2}$~s$^{-1}$).
   The curvature was modelled by using an exponential cutoff with parameters $E_{\rm f} = 14{+8\atop -4}$~keV, $E_{\rm c}=20{+6 \atop -20}$~keV (90\% c.l.). 
   The X-ray observations of \src\ showed that absorption column density ($N_{\rm H}$) in its direction displays a large variability, ranging from $\sim 10^{22}$~cm$^{-2}$ to $\sim 4.8\times 10^{23}$~cm$^{-2}$, and uncorrelated with the flux.

Shortly after the discovery of \src, \citet{Sguera06} proposed that this source was one of the so-called 
supergiant fast X-ray transient (SFXT) based mainly on the spectral properties of the hard X-ray emission and the high flux variability. 
SFXTs are a sub-class of high-mass X-ray binaries (HMXBs) hosting a compact object (likely a neutron star) that is 
accreting from the stellar wind of a massive OB supergiant. These objects show typical dynamic ranges of $\sim 10^4-10^5$ (up to $10^6$, \citealt{Romano15b})
   and sporadic flares with durations of $\sim 10^3$~s. The recorded peak X-ray luminosities during these events are usually of $10^{36}-10^{37}$~erg~s$^{-1}$ \citep{Romano15, Sidoli17, Romano23}.  
   Several accretion mechanisms have been proposed to explain the fast and strong variability of SFXTs.
   They involve gating mechanisms, settling accretion regimes, and accretion of inhomogeneous winds
   from the donor star \citep[e.g., ][]{intZand05,Grebenev07,Bozzo08,Ducci10,Shakura14}.
   
The identification of CXOU~J174042.0$-$280724 as the \chandra\ counterpart to \src\ \citep{Romano11, Heinke09}
enabled the association of the X-ray source with an infrared object \citep{Greiss11, Kaur11}. This was studied in detail by 
\citet{Greiss11} using the 4-m Visible and Infrared Survey Telescope for Astronomy (VISTA) at Paranal Observatory.
The J2000 coordinate of the optical counterpart are RA=17:40:42.0168, Dec=$-$28:07:25.050, with an astrometric fit RMS of 0.1 arcsec.
The International Astronomical Union (IAU) identifier of this object is VVV~J174042.01$-$280725.05.
Assuming extinction values based on the properties of red clump giants and the VISTA Variables in the Via Lactea (VVV) survey data \citep{Gonzalez11}, 
\citet{Greiss11} classified this star as a late type-F dwarf at $d\approx$3.8~kpc.
By exploiting archival data collected by the instrument Son OF ISAAC (SOFI) mounted on the ESO-New Technology Telescope (NTT),
\citet{Kaur11} was able to prove that VVV~J174042.01$-$280725.05 underwent a high luminosity event (of about one magnitude brighter than in the normal luminosity state)
about four days after the detection of X-ray flares caught by \swift. This was interpreted as irradiation of the
optical star by the X-rays from the compact object \citep{Romano16}. The optical data were clearly pointing
to \src\ being a low-mass X-ray binary (LMXB) rather than an HMXBs, thus excluding the previously proposed association of the source with the SFXT class. 

LMXBs are binary systems in which a neutron star (NS) or a black hole accretes matter from a low-mass donor star
with a typical mass of $\lesssim$1$M_\odot$. The accretion process occurs more commonly via an accretion disc 
but there is a number of objects where the accretion onto the compact object takes place directly from a  
stellar wind \citep[see, e.g., ][ and references therein]{Done07, Bahramian22}.
LMXBs typically have X-ray luminosities in the range $10^{33}-10^{38}$~erg~s$^{-1}$, with some sources exhibiting
outbursts that can last weeks or months and bursts on top of these, that can increase the luminosity by up to two
orders of magnitude for short periods of time ($\sim 1-100$~s).
Most LMXBs also show a dramatic X-ray spectral variability, often associated with changes in the mass accretion rate \citep[see, e.g., ][]{vanderklis06, vanParadijs98}.
\citet{Romano16} ruled out the possibility that \src\ belongs to the subclass of LMXBs called \emph{burst-only sources}\footnote{Burst-only sources show type I bursts during X-ray luminosity states that do not exceed $\sim 10^{35}-10^{36}$~erg~s$^{-1}$. This is at variance with the typical behaviour of other LMXBs, were bursts show up during persistent emission or outbursts brighter than $10^{36}$~erg~s$^{-1}$ \citep[see, e.g., ][]{Cornelisse04,Campana09}. 
Type-I bursts are rapid increases in X-ray luminosity caused by thermonuclear explosions that occur on the surface of an accreting
NSs in LMXBs. Regarding \src, the classification as burst-only source was previously proposed by \citet{Sguera06} as an alternative to the SFXT hypothesis.}
because the properties of the flares observed from \src\ substantially differ from the so-called 
type-I bursts typical of the members of this subclass. 
Based on the X-ray spectral and variability properties of \src, \citet{Romano16} also excluded that this object could be a very faint X-ray transient (VFXT), a subclass of LMXBs whose members show faint outbursts with peak
X-ray luminosities two-three orders of magnitude lower than those of other LMXBs 
(in the range $L_{\rm x}\approx 10^{34}-10^{36}$~erg~s$^{-1}$). 
About 30\% of VFXTs are also known to display type-I bursts \citep{Shaw17, King06, DelSanto10}.

In an attempt to clarify the nature of this object, in this work we report on the remarkable flaring activity detected for the first time from \src\ by \nustar\ during an observation performed in 2022.  
We also analyse the first source broadband spectrum ($\sim$0.2-60~keV) obtained by combining the \nustar\ data with a quasi-simultaneous observation carried out with \xmm\ during the persistent low luminosity state of the source. We discuss the results obtained from the X-ray data in light of the revised classification of \src\ and complement this discussion with a  comprehensive investigation of the different spectral type possibilities for the donor star hosted in this system, exploiting archival photometric data.

\section{Data analysis}
   
\subsection{NuSTAR}

\begin{figure}
   \centering
      \includegraphics[width=\columnwidth]{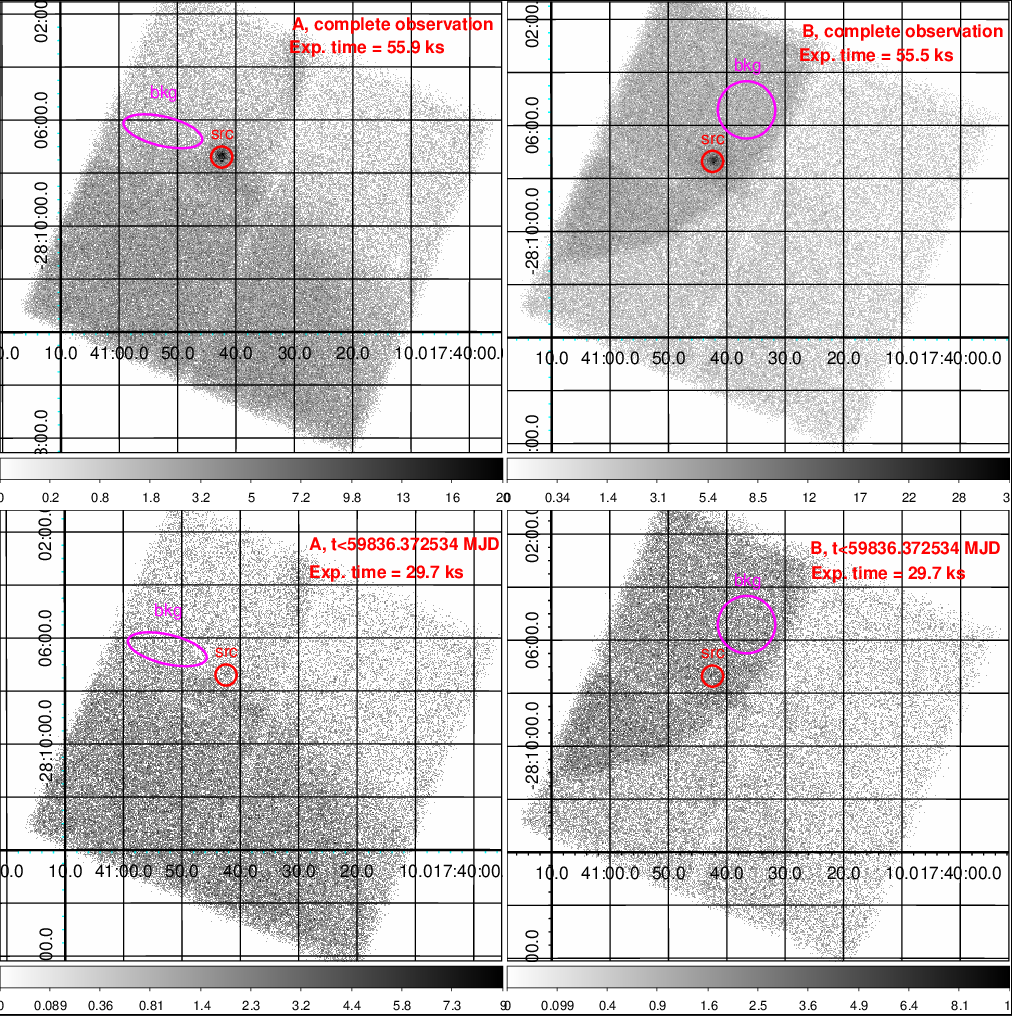}
   \caption{\nustar\ images (3-79~keV) of the \src\ field. {\it Top panels}: images for the module A (left) and B (right) obtained using the entire observation. {\it Bottom panels}: 
   images obtained during the first part of the observation ($t<59836.372534$~MJD), where \src\ is faint and its flux constant (see Fig. \ref{nustar_lcrs}). The red circle show the extraction regions centered on the target. Magenta ellipses and circles show the extraction regions used for the background. Colorbars in the bottom show the count rate for each pixel.}
   \label{nustar_straylight}
   \end{figure}

The Nuclear Spectroscopic Telescope Array (\nustar) satellite, launched on June 13, 2012,
hosts two identical co-aligned telescopes equipped with the focal plane modules FPMA and FPMB, and operates
in the 3$-$79~keV energy band \citep{Harrison13}.
\nustar\ observed \src\ on 13 and 14 September 2022, for a net exposure time of about 55~ks (see Table \ref{table log}).
We reduced the data using {\tt NUSTARDAS v2.1.2} included in
{\tt HEASOFT} v6.30.1 and the calibration files distributed with the {\tt CALDB} v20221019 \citep{Madsen22}.
For the source, we extracted events from a circular region centered on the source, 
with a radius of 24$^{\prime\prime}$ for both the FPMA and FPMB. 
We tested the effect on the final spectral results of different extraction regions for the background. The final prescription adopted to maximize the signal-to-noise ratio (S/N) of the data involved regions located on the same detector of the source (Det-0) but in a zone of the focal plane free from the emission of \src. 
As the observation was affected by stray light contaminations, we visually verified that all considered regions for the background extraction were characterised by a level of stray light contamination similar to that affecting the source. In particular, we selected for the FPMA an elliptical extraction region centered on RA: 17:40:52.49, Dec: $-$28:06:25.6 (J2000), with a major (minor) axis of $r_{\rm max}=91.283^{\prime\prime}$ ($r_{\rm min}=33.178^{\prime\prime}$) and a rotation angle of $\phi=346.79^\circ$ relative to the RA axis. For the FPMB we instead adopted a circular extraction region centered on RA: 17:40:36.61, Dec: $-$28:05:25.0, and characterised by a radius of $r=65^{\prime\prime}$ (see Fig. \ref{nustar_straylight}).

\subsection{XMM-Newton} \label{sect. xmm}

The X-ray Multi-Mirror Mission (\xmm) was launched on December 10, 1999 \citep{Jansen01}.
It hosts the European Photon Imaging Camera (EPIC) that comprises the \emph{pn}, Metal Oxide Semi-conductor 1 and 2 (MOS1 and MOS2) CCDs,
operating in the 0.2$-$12~keV energy band  \citep{Strueder01, Turner01}.
\xmm\ observed \src\ on 13 and 14 September 2022 (see Table \ref{table log}).
We reduced the data using the \xmm\ Science Analysis System (SAS v20.0.0), with the latest calibration files available in the \xmm\ calibration database (CCF).
Calibrated event lists for the \emph{pn}, MOS1, and MOS2 were obtained from the raw data files exploiting the SAS tasks {\tt epproc} and {\tt emproc}.
For the \emph{pn}, we used single- and double-pixel events, while for the MOS data we used single- to quadruple-pixel events.
We excluded time intervals were the background was too high to perform a meaningful spectral analysis using standard procedures and criteria\footnote{\url{https://www.cosmos.esa.int/web/xmm-newton/sas-thread-epic-filterbackground}}.
The net exposure time obtained for the \xmm\ observation is reported in Table \ref{table log}.
For each of the \emph{pn}, MOS1, and MOS2 cameras we extracted the source events using circular regions centered on the best known position of \src.  
The radii of these extraction regions were  $r_{pn}=11^{\prime\prime}$, $r_{\rm MOS1}=14^{\prime\prime}$, and $r_{\rm MOS2}=17^{\prime\prime}$,  respectively. 
These were derived by the SAS task {\tt eregionanalyse} to maximise the S/N. 
Background events were accumulated for each of the three cameras using extraction regions not contaminated by the emission from \src.\ The extraction regions were a circle for \emph{pn} (centered on RA:17:40:36.5703, Dec:$-$28:07:32.127 and with a radius of $r=38.5^{\prime\prime}$) and an annulus for both the MOS1 and MOS2 (the annuli were characterised by $r_{\rm in}^{MOS1}=28^{\prime\prime}$, $r_{\rm out}^{MOS1}=48^{\prime\prime}$ for the MOS1 and 
$r_{\rm in}^{MOS2}=34^{\prime\prime}$, $r_{\rm out}^{MOS2}=55^{\prime\prime}$ for the MOS2).
For a better alignment with the \nustar\ spectra, we applied corrections to the effective area of \emph{pn}, MOS1, and MOS2 spectra in accordance with the CCF Release Note {\tt XMM-CCF-REL-388}\footnote{\url{https://xmmweb.esac.esa.int/docs/documents/CAL-SRN-0388-1-4.pdf}}.

\section{Results}
\label{sect. results}

\subsection{Variability and timing analysis}

\begin{figure*}[ht!]
   \centering
   \includegraphics[height=7.43cm]{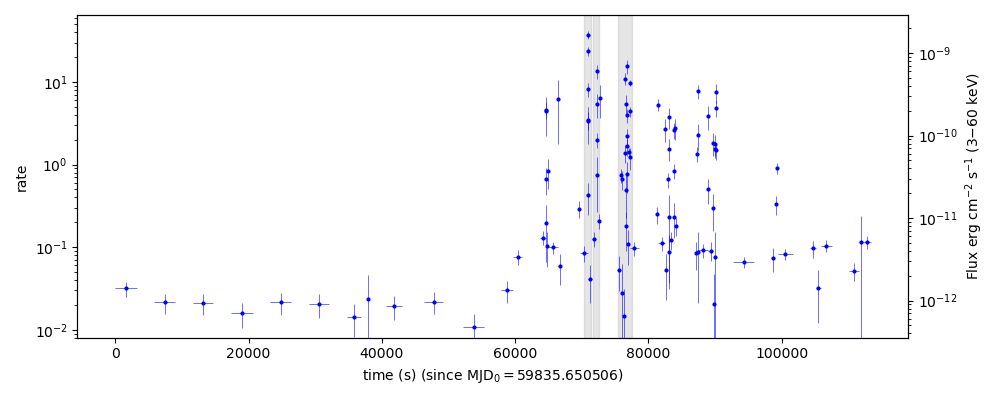}\\
 \includegraphics[height=6.5cm]{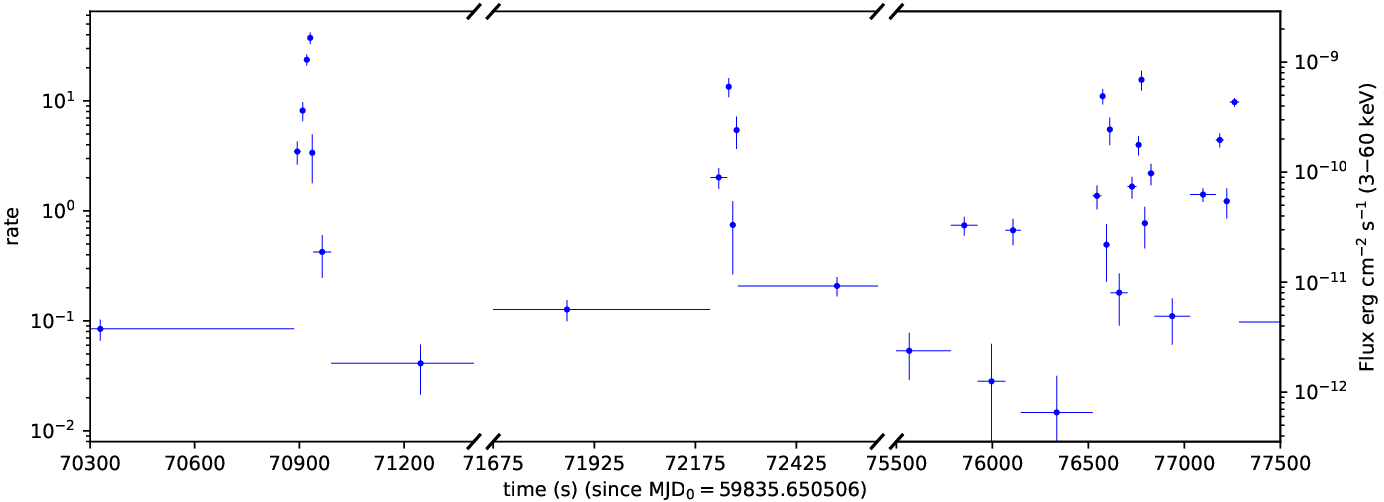}
   \caption{\emph{Top panel:} \nustar\ light curve in the energy range 3$-$60~keV obtained through Bayesian block segmentation method. \emph{Bottom panels:} three zoomed in sections of the light curve (corresponding to the three shaded gray areas in the top panel), to better show the typical time duration and structures of the flares. The ``holes'' among bins, especially noticeable in the low luminosity state of top panel, arise from the passage of \nustar\ through the South Atlantic Anomaly regions.}
   \label{nustar_lcrs}
   \end{figure*}

\begin{figure}[ht!]
   \centering
   \includegraphics[width=\columnwidth]{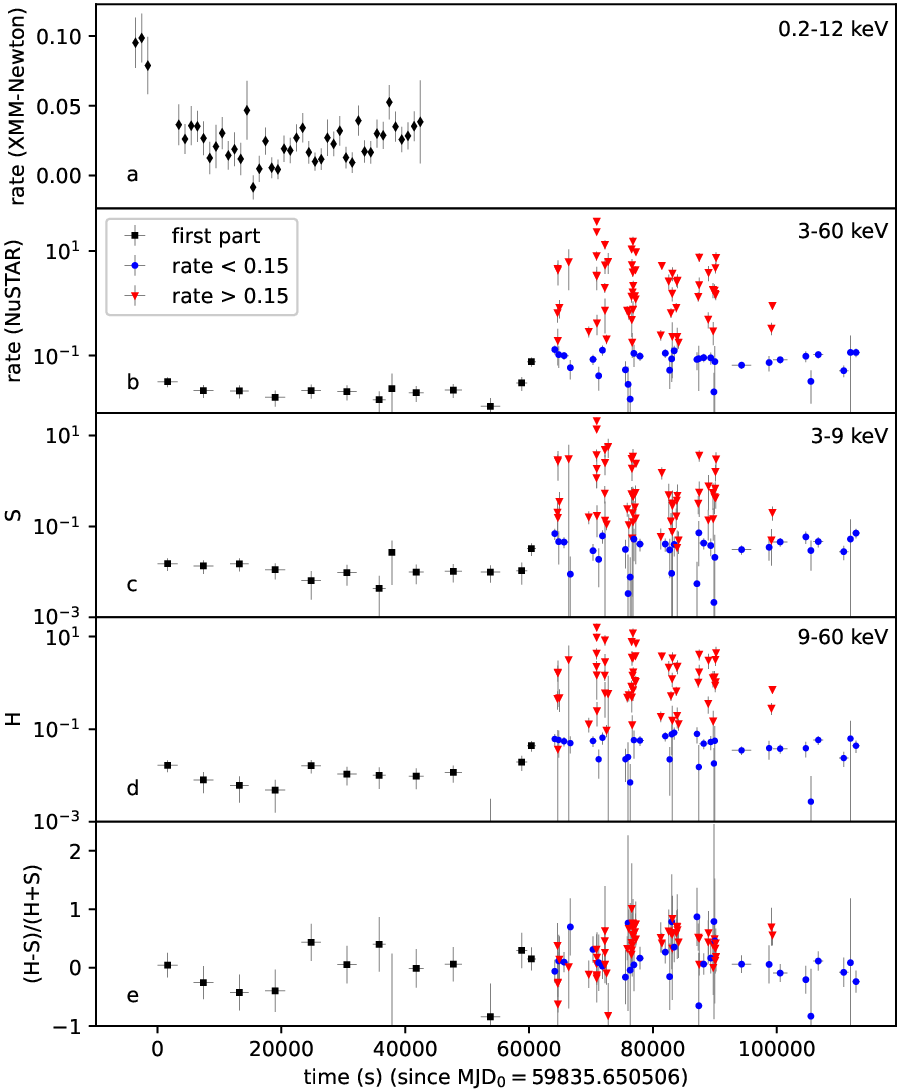}
   \includegraphics[width=\columnwidth]{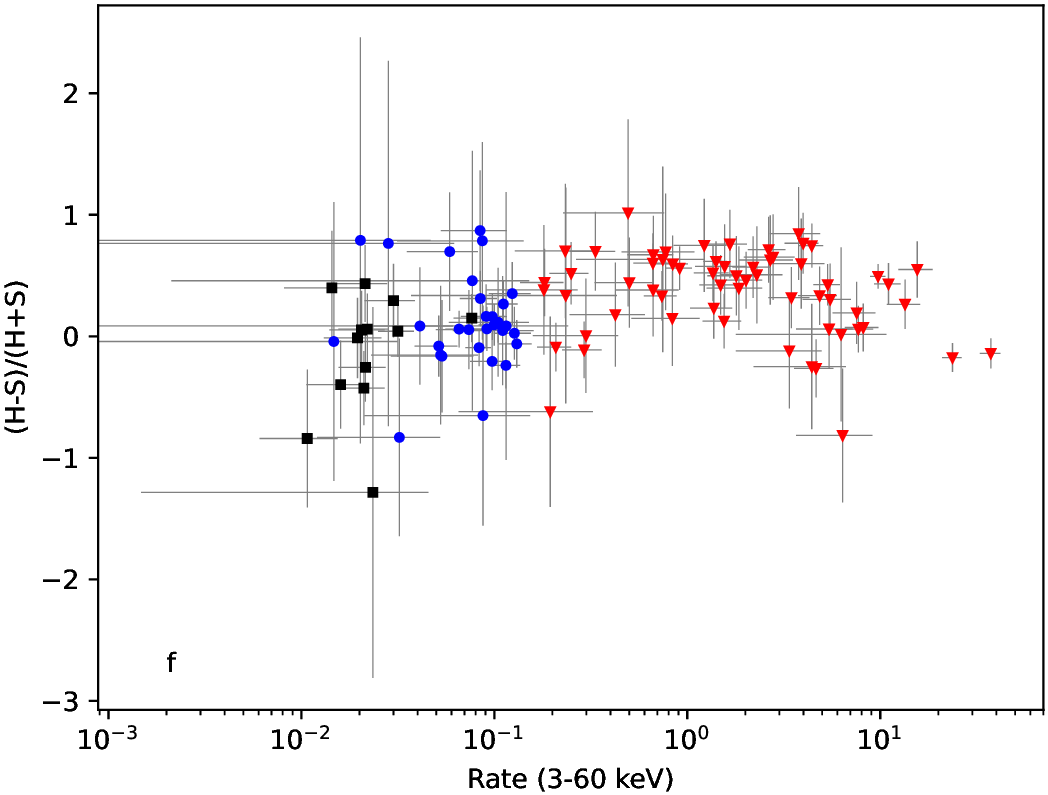}
   \caption{\emph{Panel ``a''}: \xmm/\emph{pn} light curve (0.2$-$12~keV); \emph{panel ``b''}: \nustar\  light curve (3$-$60~keV); \emph{panel ``c''}: \nustar\ light curve in the energy band 3$-$9~keV; \emph{panel ``d''}: \nustar\ light curve in the energy band $9-60$~keV; \emph{panel ``e''}: \nustar\ hardness-ratios as a function of time; \emph{panel ``f''}: \nustar\ hardness ratios as function of rate. The first part of the light curve, where the flares are absent, is displayed in all panels with black points (diamonds for \xmm, squares for \nustar). Flares: red triangles. Inter-flares: blue circles.}
   \label{nustar_HRs}
   \end{figure}

Figure \ref{nustar_lcrs} shows the background-subtracted \nustar\ light curve in the energy range 3$-$60~keV.
In the first part of the light curve ($t<59836.372534$~MJD), \src\ is barely detected in FPMB,
due to stray light contamination (see Fig. \ref{nustar_straylight}).
To increase the S/N, for the first part of the light curve we considered only data from FPMA,
less affected by the stray light.
To bring out the high dynamic range on a short scale experienced by \src, to reduce as much as possible
the bins with upper-limits, and to avoid to smear out the variability of the fast flares, the light curve is binned
using the optimal segmentation technique based on the Bayesian block representation described in \citet{Scargle13}.
The algorithm proposed by \citet{Scargle13} splits a list of of photon arrival times into an optimal maximum number of blocks
such that within each block the arrival times of the photons can be described by a Poisson distribution from a constant rate,
and adjacent blocks are statistically different.
We adopted a relatively large false positive rate probability (the probability of erroneously reporting the presence of a change point
in the data; it is used to compute the prior on the number of blocks) of $p=0.1$, to be sensitive to the fast rate fluctuations
displayed by the flares.
Once the optimal segmentation of the data was obtained, we calculated, for each bin, the rate and
its error using the standard tool for the analysis of \nustar\ data {\tt nuproducts}.
The flux conversion to obtain the right y-axis of Fig. \ref{nustar_lcrs} was obtained
adopting the spectral model of the low flux state (see Sect. \ref{sect. spectral analysis}).
Figure  \ref{nustar_lcrs} shows numerous flares, clustered in a time interval of $\sim 40$~ks,
with a dynamic range up to $\sim 2\times 10^3$.
The durations of the flares are $\sim 1-100$~s, and they can show single or multi-peak structures.
The absorbed X-ray luminosity (3$-$60~keV) during the first part of the observation, where flares were
absent (black points in Fig. \ref{nustar_HRs}) is $L_{\rm x} \approx 1.7\times 10^{33}d^2_4$~erg~s$^{-1}$
($d_4$ is the distance in units of 4~kpc).
During the second part (blue points in Fig. \ref{nustar_HRs}), the average inter-flare luminosity slightly increases
to $L_{\rm x} \approx 5\times 10^{33}d^2_4$~erg~s$^{-1}$.
The maximum luminosity reached by \src\ during the flaring activity (red points in Fig. \ref{nustar_HRs})
is $L_{\rm x} \approx 2.6\times 10^{36}d^2_4$~erg~s$^{-1}$.

   Figure \ref{nustar_HRs} shows the \nustar\ hardness ratios (HRs) as function of time and flux (panel ``e'').
   The hardness ratios are calculated as $HR = (H-S)/(H+S)$, where $S$ is the rate in the soft energy band
   $3-9$~keV, $H$ is the rate in the hard energy band $9-60$~keV. The boundary at 9~keV is chosen to ensure
   a similar number of average counts in $S$ and $H$.
   In general, Fig. \ref{nustar_HRs} does not show a dramatic HR variability,
   although a first visual inspection suggests a possible small variability in panel ``f'' (i.e., HRs vs flux).
   A more thorough exploration of the spectral variability is presented in Sect. \ref{sect. spectral analysis}.

   Panel ``a'' of Fig. \ref{nustar_HRs} shows the background-subtracted \xmm/\emph{pn} light curve in the energy range 0.2$-$12~keV, with a binsize of 1~ks.
   The HRs based on the \xmm\ data, in the energy bands 0.2-3~keV and 3-12~keV, do not show significant variability in hardness.

   We searched for periodic signals in the 0.3$-$12~keV \emph{pn} events corrected to the solar system barycenter
   with the {\tt barycen} SAS task and in the 3$-$60~keV \nustar\ events corrected to the solar system barycenter
   with the {\tt barycorr} task from {\tt ftools}, using a Rayleigh test $Z^2$ (see, e.g., \citealt{Buccheri83}) from 1 to 3 harmonics.
   No statistically significant pulsations were detected in the frequency range $0.001-175.44$~Hz for \xmm/\emph{pn}
   and $0.001-1000$~Hz for \nustar. We used the approach described in \citet{Brazier94} to calculate the
   3$\sigma$ upper limit on the pulsed fraction $p_{\rm f}$ of a sinusoidal signal. We found, for \xmm/\emph{pn}, $p_{\rm f}=50$\%,
   and for \nustar, $p_{\rm f}=40$\%.

   \subsection{Spectral analysis} \label{sect. spectral analysis}

\begin{figure}[ht!]
   \centering
    \includegraphics[width=\columnwidth]{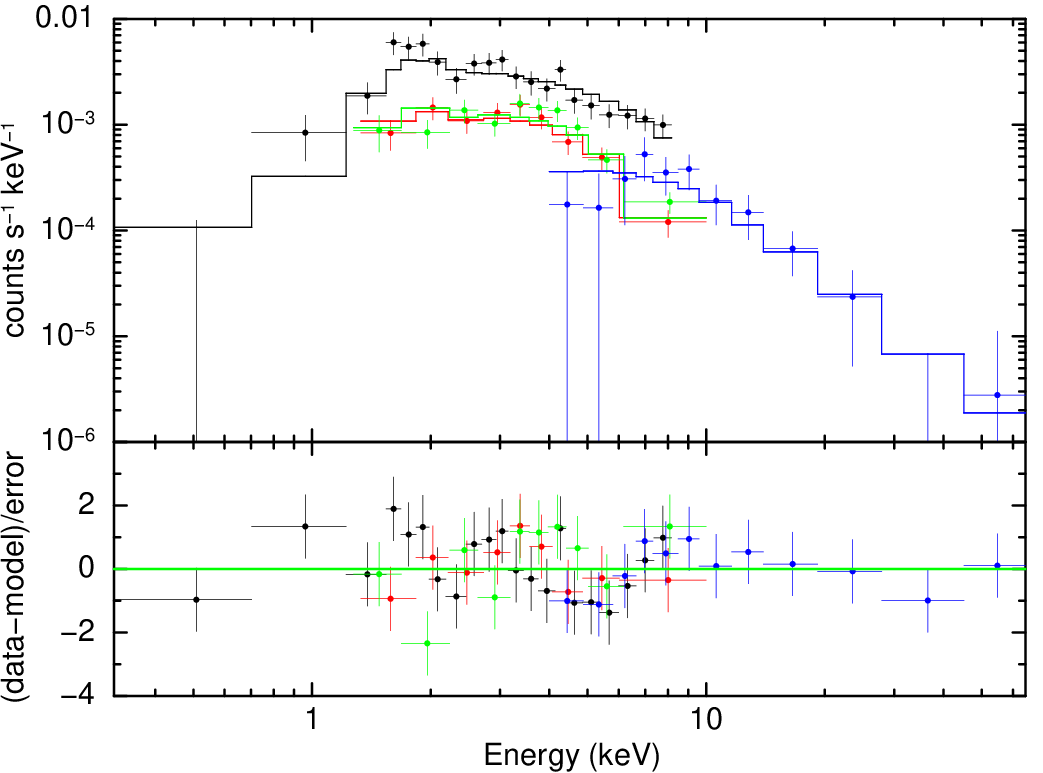}
   \caption{\xmm\ (black: \emph{pn}; red: MOS1; green: MOS2) and \nustar\ (blue: module A) spectra of \src\
     during the first part of the light curve (black points in Fig. \ref{nustar_HRs}), fitted with an
   absorbed power law (see Table \ref{Table nustar_spectra}). The lower panel shows the residuals of the fit.}
   \label{xmm_nustar_spectra}
   \end{figure}

\begin{figure}[ht!]
   \centering
      \includegraphics[width=\columnwidth]{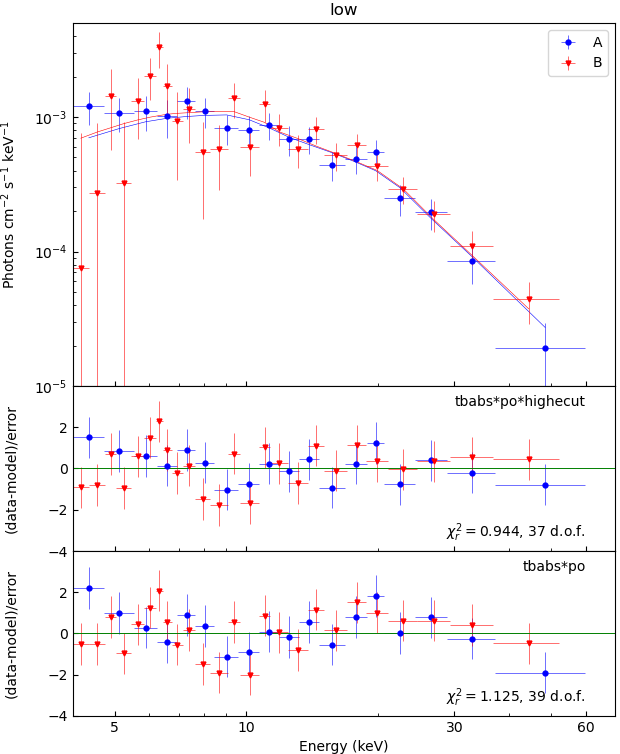}
   \caption{\nustar\ (blue circles: module A; red triangles: module B) spectra of \src\ during the low luminosity state between flares (blue circles in Fig. \ref{nustar_HRs}). \emph{Top panel:} the spectra are fitted with an absorbed power law with high-energy cutoff. \emph{Middle panel:} residuals of the same fit of the top panel. \emph{Bottom panel:} residuals of the fit of the spectra with an absorbed power law. See Table \ref{Table nustar_spectra} for the best fit parameters.}
   \label{nustar_spectra_low}
   \end{figure}

\begin{figure}[ht!]
   \centering
      \includegraphics[width=\columnwidth]{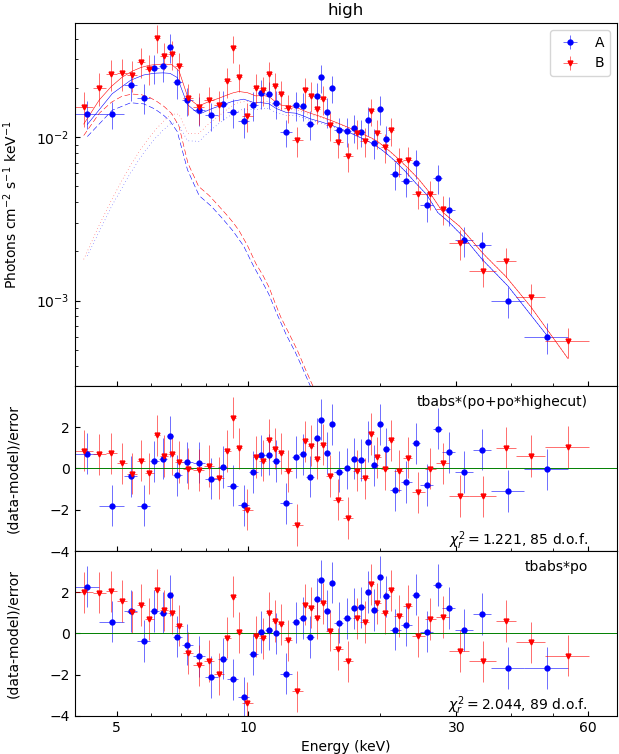}
   \caption{\nustar\ (blue circles: module A; red triangles: module B) spectra of \src\ during the high luminosity state (i.e. flares; red triangles in Fig. \ref{nustar_HRs}). \emph{Top panel:} the spectra are fitted with two power law, both absorbed, one of them  with high-energy cutoff. \emph{Middle panel:} residuals of the same fit of the top panel. \emph{Bottom panel:} residuals of the fit of the spectra with an absorbed power law. The purpose of this panel is to highlight the presence of a feature in the residuals at $\sim 8.6$~keV, which requires the use of a more complex model, as described in Sect. \ref{sect. spectral analysis}. See Table \ref{Table nustar_spectra} for the best fit parameters, also with other spectral models.}
   \label{nustar_spectra_low_high}
   \end{figure}

   We divided the data in three subsets to search for weak spectral variability
   not detectable with the hardness ratios.
   We divided the data using the same scheme shown in Fig. \ref{nustar_HRs}:
   \begin{itemize}
     \item \emph{1) first part} ($t<59836.372534$~MJD), where the flux is low and flares are absent;
       These data were fitted simultaneously with the \xmm\ data. (black points in Fig. \ref{nustar_HRs});
     \item \emph{2) low}: inter-flares emission (rate<0.15) observed by \nustar\ after $t>59836.372534$~MJD (blue points in Fig. \ref{nustar_HRs});
     \item \emph{3) high}: flares (rate>0.15) observed by \nustar\ after $t>59836.372534$~MJD (red points in Fig. \ref{nustar_HRs}).
   \end{itemize}
   \nustar\ and \xmm\ spectra were rebinned so as to have at least 25 counts per bin to enable the use
   of $\chi^2$ statistic as a fit statistic. 
   Renormalization constant factors were included in the spectral fitting to account for intercalibration uncertainties between instruments.
   We used the {\tt tbabs} model and the interstellar medium abundances {\tt wilm} in {\tt XSPEC}\footnote{{\tt XSPEC} version 12.12.1c; \citep{Arnaud96}.}
   to model the photoelectric absorption \citep{Wilms00}.
   Errors on spectral fit parameters indicate 1$\sigma$ confidence level throughout the paper.
   We fitted the spectra with different models.
   In the following, we concentrate on the simplest phenomenological models that provide a reasonably good fit.
   The best-fit models and parameters are reported in Table \ref{Table nustar_spectra}.
   The corresponding spectra and residuals are shown in Figures 
   \ref{xmm_nustar_spectra}, \ref{nustar_spectra_low}, and \ref{nustar_spectra_low_high}.
   For the \emph{first part}, we obtained a good fit with an absorbed power law ($\chi^2_\nu=0.963$, 46 d.o.f.).
   For the \emph{low} state, the same model gives an acceptable fit ($\chi^2_\nu=1.125$, 39 d.o.f.),
   although we obtained a slightly better fit with an absorbed power law with high-energy cutoff
   ($\chi^2_\nu=0.944$, 37 d.o.f.).
   For the \emph{high} state, to model residuals at $\approx 8.5$~keV (see bottom panel of
   Fig. \ref{nustar_spectra_low_high}), it is necessary to employ a more complex model.
   We obtained the best fits with an absorbed power law with a
   Gaussian in absorption $E_{\rm gabs}=8.6{+0.6\atop -0.5}$~keV ($\chi^2_\nu=1.240$, 86 d.o.f.),
   or a power law with high energy cutoff plus a steep power law, both components
   absorbed ($\chi^2_\nu=1.221$, 85 d.o.f.).
   For the \emph{high} state, we evaluated the chance probability of improvement of the fit by adding
   the high energy cutoff, simulating $10^4$ data sets with the 
   {\tt simftest} routine of {\tt XSPEC}. We find that the probability that data are consistent
   with a model without the high energy cutoff component is 0.30\%. 
   The best fit model obtained for the \emph{high} state data applied to the \emph{low} 
   state\footnote{All spectral parameters frozen, except for the normalization of the first power law, and the ratio of the normalization of the second power law with respect to the first one.} gives $\chi^2_\nu=0.773$, 41 d.o.f. Therefore, the relatively low statistic of the \emph{low} state spectrum does not allow to determine if there is a spectral variability between these two flux states. A similar fit applied to the \emph{first part} of the data gives an unacceptable fit, even if the column density is free ($\chi^2_\nu=8.568$, 48 d.o.f.).
   Vice versa, the best fit model obtained for the \emph{first part} of the data applied to the \emph{low} state\footnote{All spectral parameters frozen, except for the normalization of the power law.} gives an acceptable fit only if the column density is free ($\chi^2_\nu=1.248$, 40 d.o.f.).
   However, the absorption column density we obtain, $N_{\rm H}\approx 9\times 10^{23}$~cm$^{-2}$, seems too high, compared to the typical values in this source and other typical X-ray binaries. A similar fit applied to the \emph{high} state gives an unacceptable $\chi^2_\nu=2.629$, 90 d.o.f., and a very high $N_{\rm H}\approx 10^{24}$~cm$^{-2}$.
   Therefore, it seems that there is a significant spectral variability, especially driven by the increase of the column density, between the first and second (low and high) part.
   
   The low state spectrum from module B shows an emission feature at $\sim6.4$~keV, which could be an iron line (Fig. \ref{nustar_spectra_low}). To understand if it is necessary to add a Gaussian component to model it, we rebinned the spectra of module A and B to have a minimum of 1 count per bin and we used the
   W statistic \citep{Wachter79, Cash79}  to find the best fit. Then, we evaluated the chance probability of improvement of the fit by adding this component simulating $10^4$ data sets with the {\tt simftest} routine of {\tt XSPEC}. We find that the probability that the data are consistent with a model without the Gaussian component is 0.13\,\%. Therefore, the hypothesis of the absence of an iron line cannot be rejected at a 4$\sigma$ confidence level.

\begin{table}
\begin{center}
\caption{Best-fit spectral parameters to describe the \emph{first part} (black points in Fig. \ref{nustar_HRs}), \emph{low} luminosity state (blue points in Fig. \ref{nustar_HRs}), and \emph{high} luminosity state (i.e., flares; red points in Fig. \ref{nustar_HRs}).}
\vspace{-0.3cm}
\label{Table nustar_spectra}
\resizebox{\columnwidth}{!}{
\begin{tabular}{lccc}
\hline
\hline
\noalign{\smallskip}
                                   & \multicolumn{2}{c}{First part} \\
\noalign{\smallskip}
\hline
\noalign{\smallskip}
Parameters                         &         Model A$^a$             &                  &                   \\
\noalign{\smallskip}
\hline
\noalign{\smallskip}
$N_{\rm H}$ ($10^{22}$~cm$^{-2}$)     &      $3.32{+0.56 \atop -0.51}$  &                  &                    \\
\noalign{\smallskip}
$\Gamma$                           &   $1.47{+0.17 \atop -0.17}$     &                  &                    \\
\noalign{\smallskip}
norm$_{\rm pow}$                     & $4.1{+1.4 \atop -1.0}\times 10^{-5}$ &              &                    \\
\noalign{\smallskip}
constant (MOS1 wrt pn)            &    $0.86{+0.09 \atop -0.08}$    &                   &                     \\
\noalign{\smallskip}
constant (MOS2 wrt pn)            &    $0.83{+0.09 \atop -0.08}$    &                   &                     \\
\noalign{\smallskip}
constant (NuSTAR-A wrt pn)        &    $1.18{+0.25 \atop -0.23}$    &                   &                     \\
\noalign{\smallskip}
$\chi^2$ (d.o.f.)                  &       44.31 (46)                &                  &                     \\
\noalign{\smallskip}
Null hypothesis prob.  &  0.543              &                  &                     \\
$F_{\rm x, 0.3-60~keV}$ (erg~cm$^{-2}$~s$^{-1}$) & $8.9\pm1.8\times 10^{-13}$  &                  &                      \\
$F_{\rm x, 4-60~keV}$ (erg~cm$^{-2}$~s$^{-1}$) & $8.3\pm1.8\times 10^{-13}$  &                  &                      \\
\noalign{\smallskip}
\hline
\noalign{\smallskip}
                                   & \multicolumn{2}{c}{low} \\
\noalign{\smallskip}
\hline
\noalign{\smallskip}
Parameters                         &         Model A$^a$             &      Model B$^b$                     &           \\
\noalign{\smallskip}
\hline
\noalign{\smallskip}
$N_{\rm H}$ ($10^{22}$~cm$^{-2}$)     &      $39{+22 \atop -18}$        &    $<48$ (3$\sigma$)                 &           \\
\noalign{\smallskip}
$\Gamma$                           &   $0.96{+0.19 \atop -0.17}$     &  $0.25{+0.30 \atop -0.14}$           &           \\
\noalign{\smallskip}
norm$_{\rm pow}$                      & $9.4{+7.7 \atop -3.9}\times 10^{-5}$ &  $1.34{+1.70 \atop -0.41}\times 10^{-5}$ &   \\
\noalign{\smallskip}
$E_{\rm cut}$ (keV)                  &                                 &  $20.4{+10.2 \atop -2.4}$            &           \\
\noalign{\smallskip}
$E_{\rm fold}$ (keV)                 &                                 &  $22.68{+7.6 \atop -6.1}$            &           \\
\noalign{\smallskip}
constant (B wrt A)                 &    $1.11{+0.11 \atop -0.10}$    & $1.08{+0.11 \atop -0.10}$            &           \\
\noalign{\smallskip}
$\chi^2$ (d.o.f.)                  &       43.87 (39)                & 34.92 (37)                           &            \\
\noalign{\smallskip}
Null hypothesis prob.    &   0.273         &    0.567          &                     \\
$F_{\rm x, 4-60~keV}$ (erg~cm$^{-2}$~s$^{-1}$) & $8.8\pm 0.9\times 10^{-12}$ & $7.8\pm 0.7\times 10^{-12}$  &            \\
\noalign{\smallskip}
\hline
\noalign{\smallskip}
                                   & \multicolumn{2}{c}{high} \\
\noalign{\smallskip}
\hline
\noalign{\smallskip}
Parameters                         &      Model C$^c$                &      Model D$^d$                     &       Model E$^e$               \\
\noalign{\smallskip}
\hline
\noalign{\smallskip}
$N_{\rm H}$ ($10^{22}$~cm$^{-2}$)     &   $30.6{+6.0 \atop -5.6}$     &   $121{+13 \atop -13}$              &    $85{+14\atop -15}$        \\
\noalign{\smallskip}
$\Gamma_1$                           &   $1.56{+0.21 \atop -0.18}$     &   $1.20{+0.09 \atop -0.09}$      &    $0.45{+0.23\atop -0.30}$     \\
\noalign{\smallskip}
norm$_{\rm pow 1}$                  & $1.6{+1.7 \atop -0.8}\times 10^{-2}$ & $4.93{+1.7 \atop -1.2}\times 10^{-3}$ &  $5.8{+5.8\atop -3.0}\times 10^{-4}$  \\
\noalign{\smallskip}
$E_{\rm cut}$ (keV)                  &                                 &                                      &   $19.6{+1.7 \atop -1.5}$   \\
\noalign{\smallskip}
$E_{\rm fold}$ (keV)                 &                                  &                                     &   $28.8{+8.2 \atop -7.1}$    \\
\noalign{\smallskip}
$E_{\rm gabs}$ (keV)                 &     $8.6{+0.6 \atop -0.5}$      &                                      &                                 \\
\noalign{\smallskip}
$\sigma_{\rm gabs}$ (keV)            &     $4.5{+1.0 \atop -0.9}$      &                                      &                                 \\
\noalign{\smallskip}
Strength$_{\rm gabs}$                &       $11{+6 \atop - 4}$        &                                      &                                 \\
\noalign{\smallskip}
$\Gamma_2$                           &                               &   $8.6{+0.9 \atop -0.8}$          &    $5.6{+1.0\atop -0.9}$     \\
\noalign{\smallskip}
norm$_{\rm pow 2}$                  &                                   & $3.25{+14.7 \atop -3.1}\times 10^{3}$ &  $12{+71\atop -7}$  \\
\noalign{\smallskip}
constant (B wrt A)                 &    $1.14{+0.05 \atop -0.05}$    &  $1.15{+0.05 \atop -0.05}$            & $1.15{+0.05\atop -0.05}$       \\
\noalign{\smallskip}
$\chi^2$ (d.o.f.)                  &    106.62 (86)                   &   116.91 (87)                        &   103.79 (85)                   \\
\noalign{\smallskip}
Null hypothesis prob.             &   0.0653           &   0.0180         &    0.0812              \\
$F_{\rm x, 4-60~keV}$ (erg~cm$^{-2}$~s$^{-1}$) & $1.81\pm0.08\times 10^{-10}$ & $1.89\pm0.09\times 10^{-10}$ & $1.79\pm0.09\times 10^{-10}$  \\
\noalign{\smallskip}
\hline
\end{tabular}
}
\end{center}
    {\small Notes. Errors are at 1$\sigma$ confidence level. Fluxes are absorbed.
       $^a$: {\tt const*tbabs*po};
       $^b$: {\tt const*tbabs*po*highecut};
       $^c$: {\tt const*tbabs*po*gabs};
       $^d$: {\tt const*tbabs*(po + po)};
       $^e$: {\tt const*tbabs*(po*highecut +po)}.
       }
\end{table}

   \section{Optical counterpart}
   \label{Sect. Optical counterpart}

\begin{figure*}
  \centering
  \def\big{\includegraphics[height=12cm]{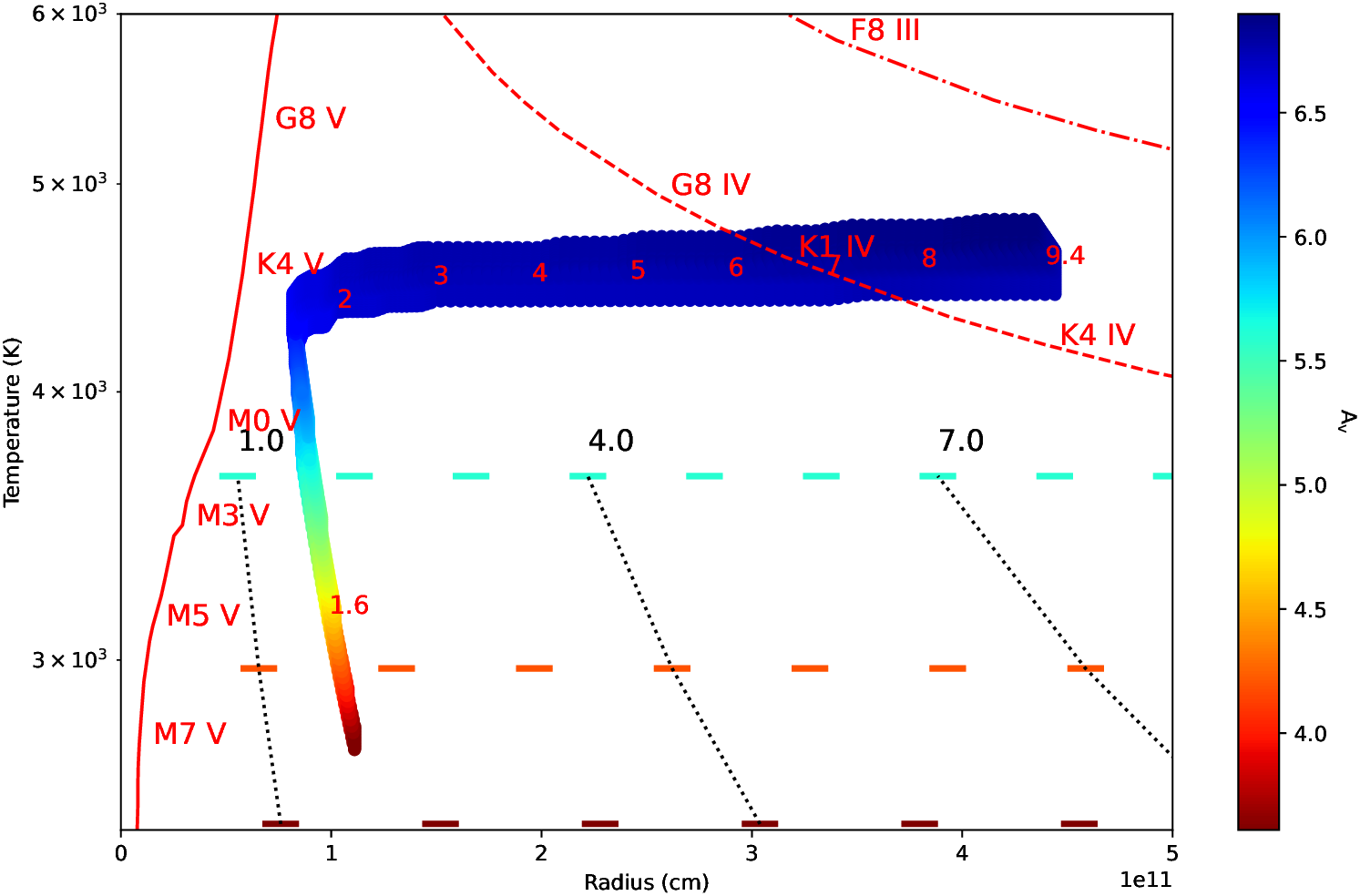}}
\def\little{\includegraphics[height=5.0cm]{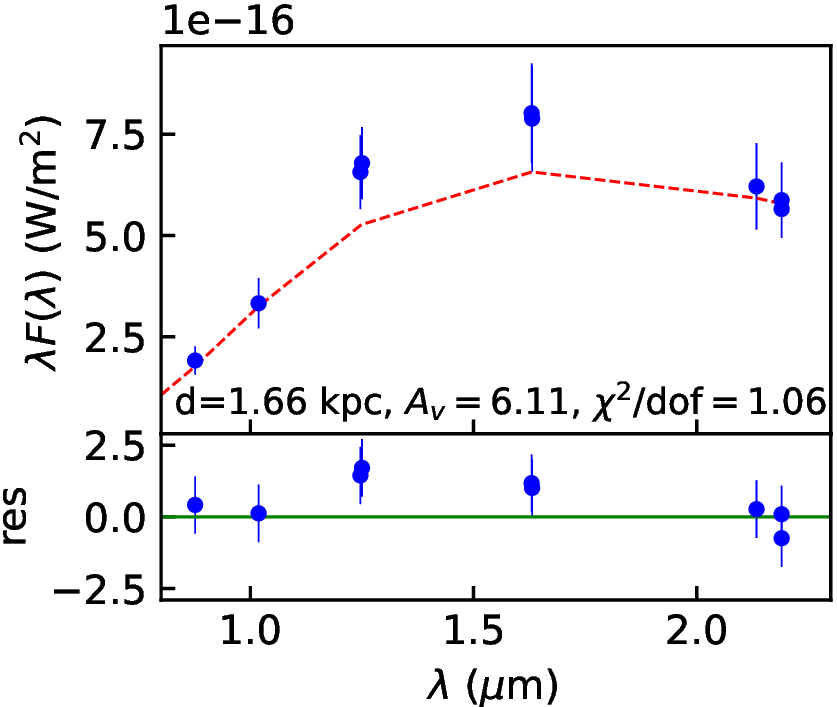}}
\stackinset{c}{90pt}{b}{38pt}{\little}{\big}
\caption{Best values of effective temperature, radius, and extinction, obtained from the fit of the photometric SED of \src\ (see Table \ref{Table sed}) with an absorbed blackbody model, with two different methods. \emph{Method one:} best fit values of $T$ and $R$ from the fit of the photometric SED, assuming distances in the range 1-30~kpc and for three values of $A_{\rm v}$ (see main text for more details). These solutions show up as three horizontal lines, whose colours reflect the values assumed for $A_{\rm v}$ (see vertical colour bar on the right). A grid of distances (in units of kpc) in black colour is overplotted. \emph{Method two:} The ``$\Gamma$-shape'' coloured area shows the best fit parameters ($\chi^2_{\rm red}<2$) $T$ and $R$ for different values of $d$, where $A_{\rm v}$ dependency on the distance is based on \citet{Green19}.
For this method, the distances (in units of kpc) from 1.6 to 9.4~kpc are overplotted using red colours. 
The most relevant spectral classes for our analysis are shown in red solid (main sequence), dashed (sub-giant), and dot-dashed (giant) lines.
The inset figure shows an example of best-fitting SED (dashed red line). Blue points are the photometric measurements (see Table \ref{Table sed}). The fit residuals (observed-model)/error are shown in the lower panel.}
\label{optical counterpart}
\end{figure*}

   We assessed the nature of the donor star in \src\ by using photometric measurements
   of VVV~J174042.01$-$280725.05, based on
   the Panoramic Survey Telescope and Rapid Response System (Pan-STARRS) catalogue \citep{Chambers16} and on
   the VVV Survey. The survey consists of images in five bands
   obtained by the VISTA InfraRed CAMera (VIRCAM) that equips the VISTA,
   a 4\,m telescope specialized in wide field surveys. The photometric information used to build the Spectral Energy Distribution (SED)
   of VVV~J174042.01$-$280725.05 is reported in Table \ref{Table sed}.
   The star is very faint and no \emph{Gaia} DR3 counterpart is catalogued within the VISTA $3\sigma$ error region \citep{Greiss11, Vallenari22}.   
   Given the small number of points available in the SED, we fitted it with a simple absorbed blackbody model, which is
   in first approximation a reasonable model to describe the optical/near-infrared spectrum from a star:
   \begin{equation} \label{eq. bb}
     \lambda F(\lambda) = \frac{2\pi hc^2}{\lambda^4}\left[ \frac{1}{e^{hc/(\lambda kT)}-1} \right] \frac{R^2}{d^2} 10^{-0.4A_\lambda} \mbox{ .}
   \end{equation}
   $T$ is the effective temperature of the star, $R$ its radius, $d$ is the distance from the Sun, and $A_\lambda$ is the absorption
   for the wavelength $\lambda$.
   We calculated $A_\lambda$ for each $\lambda$ reported\footnote{taken from the VizieR Photometry viewer: \url{http://vizier.cds.unistra.fr/vizier/sed/}} in Table \ref{Table sed} by using the analytical expression
   of \citet{Cardelli89} and \citet{ODonnell94}, and assuming  $R_{\rm v}=3.1$.
   A close look at Eq. \ref{eq. bb} shows a degeneracy between parameters $R$ and $d$. In addition, due to the limited number of points of our SED,
   there is also some degeneracy between $A_{\rm v}$ and $T$. Therefore, only two parameters out of four can be constrained simultaneously.
   To overcome these limitations, we have taken two approaches.
   First, we fitted the SED with Eq. \ref{eq. bb} for assumed distances in the range 1-30~kpc, with steps of 1~kpc,
   and we also fixed the extinction $A_{\rm v}$ assuming the lowest absorption column density value.
   That was measured in a \xmm\ observation carried out on 6 March 2016 (obsid: 0764191301).
   The value of $N_{\rm H}$ reported in \citep{Romano16} ($N_{\rm H}=0.77{+0.70\atop-0.48}\times 10^{22}$~cm$^{-2}$, 90\% c.l. unertainties) was based on the use of {\tt phabs} absorption model in {\tt XSPEC}, with the abundance of elements by \citet{Anders89}. In environments with high column density, these choices for the absorption model and abundances could give significantly different results compared to using the {\tt tbabs} model with {\tt wilm} abundances. Therefore, we re-analysed these data, using SAS 20.0, and {\tt tbabs} with {\tt wilm} abundances, coherently with our analysis method described in Sections \ref{sect. xmm} and \ref{sect. spectral analysis}. We obtain $N_{\rm H}=1.2\pm0.4\times 10^{22}$~cm$^{-2}$ (1$\sigma$ c.l.).
   We converted $N_{\rm H}$ to $A_{\rm v}$ using the relation given in \citet{Foight16}.
   The best fit parameters $T$ and $R$ are displayed in Fig. \ref{optical counterpart}: they consist of three horizontal lines, for the mean, upper, and lower limits
   of $A_{\rm v}$ (given the uncertainties on $N_{\rm H}$). A grid of distances in black colours is overplotted. The colours of these horizontal lines follow the scheme of the $A_{\rm v}$ colour bar.
   The best-fit values of $T$ and $R$ obtained with this method have $\chi^2_{\rm red}$
   ranging from 1.19 to 2.18, with $\chi^2_{\rm red}=1.63$ (8 d.o.f.) for the solutions corresponding 
   to $A_{\rm v}=4.2$ ($N_{\rm H}=1.2\times 10^{22}$~cm$^{-2}$; horizontal line with orange colour in Fig. \ref{optical counterpart}).
   This method does not take into account that $A_{\rm v}$ increases with $d$ and, consequently, solutions for many of the assumed distances are obviously wrong.
   Therefore, we adopted a second approach, in which $A_{\rm v}$ is a function of $d$. 
   We used the directional 3D maps of interstellar dust reddening and extinction {\tt bayestar2019}, based on \emph{Gaia}, Pan-STARRS~1, and Two Micron All Sky Survey (2MASS) \citep{Green19}. Reddening were converted to $A_{\rm v}$ using the 
   appropriate conversion described in \citet{Green19}\footnote{see also: \url{http://argonaut.skymaps.info/usage}},
   and adopting the 5-95\% percentile boundaries. The range of reliable distances of these maps in the direction of our target is $1.59-9.40$~kpc\footnote{Obtained using {\tt dustmaps}, \citep{Green18}.}.
   The best fit parameters $T$ and $R$ for different values of $d$ and $A_{\rm v}(d)$ are shown in Fig. \ref{optical counterpart} (``$\Gamma$'' shape).
   The distances are overplotted in red colours. Figure \ref{optical counterpart} also shows the most relevant spectral classes for our analysis, taken from \citet{deJager87} and \citet{Pecaut13}. 
   All the best fit parameters obtained with the second method and shown in Fig. \ref{optical counterpart} are from fits that give $\chi^2_{\rm red} < 2$.
   We obtain the best fits ($\chi^2_{\rm red}=1.06$), 
   for a star at a distance $\sim 1.66$~kpc, with temperature $T\approx 4\times 10^3$~K and radius $R\approx 9\times 10^{10}$~cm. 
   These values of $T$ and $R$ and those obtained assuming $A_{\rm v}=4.2\pm 1.4$ (based on the lowest $N_{\rm H}=(1.2\pm 0.4 )\times 10^{22}$~cm$^{-2}$ from X-ray observations)
   correspond to a quite peculiar K7-M7~sub-subgiant type star. 
   The coloured area in Fig. \ref{optical counterpart} shows that other classifications are possible, indicating that the spectral type of the donor star of \src\ is poorly constrained.
   It is important to point out that the spectral classification study presented here is to be taken with due caution. We recall that the near-infrared variability observed by \citet{Kaur11} could indicate that at least in some occasions (when the source is brighter) there is a significant contribution to the observed near-infrared emission produced by the reprocessing of the X-ray emission from the compact object on an accretion disc (whose presence is not certain) or on the surface of the companion star \citep{Romano16}. 
   If this is the case, considering the magnitude values when the system is faint should have minimised the impact of reprocessing on the optical near-infrared emission. Nevertheless, should this still be present, it would  mean that the donor star of \src\ is even weaker than has been measured, further supporting the LMXB scenario with a particularly light and small donor star. Due to the lack of measurements of the fundamental parameters of the system, above all its distance from the Sun and the orbital separation, it is not possible to unambiguously quantify the contribution of reprocessing to the observed emission, and therefore it is not possible to determine whether this significantly affects the result reported in this paper.

\begin{table}
\begin{center}
\caption{List of the energy fluxes of the optical counterpart of \src. We selected sources within 3$\sigma$ radius centered on the position of VVV~J174042.01$-$280725.05.}
\vspace{-0.3cm}
\label{Table sed}
\resizebox{\columnwidth}{!}{
\begin{tabular}{lcccll}
\hline
\hline
\noalign{\smallskip}
   wavelength &       flux        &        eflux        &    filter   &    catalogue  & Reference$^a$ \\
    $\mu m$   &  $10^{-6}$ (Jy)    &   $10^{-6}$ (Jy)     &             &             &               \\
\hline
\noalign{\smallskip}
 0.748   & 14.2     &  0.7$^b$  & i & Pan-STARRS & 1 \\
   0.876 &   55.9   &	10.2 &	Z & VISTA  & 2  \\
   1.02  &   113    &	21   &	Y & VISTA  & 2  \\
   1.25  &   273    &	38   &	J & VISTA  & 2  \\
   1.25   &	283  &	37  &	J & VVV    & 3  \\
   1.63	 &   429    &	71   &	H & VISTA & 2   \\
   1.63   &	436  &	67  &	H & VVV       & 3  \\
   2.13  &   442    &	76   &	Ks& VISTA  & 2  \\
    2.19  &	413  &	13  &	K &  VIRAC    & 4 \\
   2.19   &	429  &	68  & 	K & VVV     & 3  \\
\noalign{\smallskip}
\hline
\end{tabular}
}
\end{center}
\footnotesize Notes: $^a$: References. (1) \citet{Chambers16}; (2) \citet{Minniti17}; (3) \citet{Herpich21}; (4) \citet{Smith18}.
$^b$: including systematic errors for sources inside the Galactic plane \citep{Magnier20}.
\end{table}

   \section{Discussion}

\nustar\ detected \src\ in a flaring state characterised by a variability
   as large as three orders of magnitude on time scales of a few tens of seconds.
   If we consider the lowest and highest fluxes measured for this source since its discovery,
   they span more than four orders of magnitude.

In Sect. \ref{sect. results} we showed that one of the possible spectral
models adopted to describe the source emission during the flaring period is a power law subjected to a large extinction at the lower energies ($\lesssim$3~keV) due to a high column density and featuring a Gaussian absorption feature centered at $E_{\rm gabs}\approx 8.6$~keV. 
Similar absorption features are often detected in the X-ray spectra of accreting NSs and widely interpreted as cyclotron resonant scattering features (CRSFs; also collectively termed ``cyclotron lines''). CRSFs provide a direct measurement of the NS magnetic field strength close to the surface of the compact object through the simple relation $E_{\rm cyc} \approx 11.6B_{12}$~keV \citep[see, e.g.,][for a recent review]{Staubert19}. We would thus obtain for the case of \src\ a magnetic field of $B\approx 7\times 10^{11}$~G. If confirmed by future observations, this would imply that \src\ hosts a NS with a relatively strong magnetic field, uncommon in LMXBs \citep[but see, e.g., the cases of GRO\,J1744-28 and 4U\,1822-371 for examples of strongly magnetized NSs in LMXBs, as well as the cases of IGR\,J17329-2731 and 4U\,1700+24 for strongly magnetized NSs in SyXBs;][]{gro,4u1,4u2,symb1,symb2}. Note that, as discussed in Sect.~\ref{sect. spectral analysis}, the available \nustar\ data on \src\ can also be successfully fit with alternative models that do not comprise a CRSF. Therefore, the presence of this absorption component and the derivation of a putative NS magnetic field estimate have to be taken with caution (and possibly confirmed by future observations). 
The spectral analysis shows also a weak hardening and increase of $N_{\rm H}$
between the first part of the observation, where the flaring activity is absent, and the second part.
The increase in the absorption might reflect an increase of the gravitationally captured mass by the compact object.
The hardening could be ascribed to a more efficient Inverse Compton scattering of the soft X-ray photons emitted in the vicinity 
of the compact object by the increasing number of accreting electrons.

   Our current limited knowledge on fundamental characteristics of the stellar components hosted in \src,\ such as the orbital period, 
   distance between the two stars, eventual spin period and magnetic field strength of the compact object, prevents us from adopting  a quantitative approach
   to determine the real nature of this source and the accretion mechanisms triggering its X-ray variability. 
   From the X-ray spectral point of view, the \nustar\ and \xmm\ data would be qualitatively compatible with what is expected from accreting pulsars 
   with high magnetic field ($B \gtrsim 10^{12}$~G; e.g.\ \citealt{Kretschmar19}). 
   In the following, we thus discuss the similarities and differences between the properties of the  X-ray variability shown by \src\
   with those from the other binary systems, specifically those hosting strongly magnetized NSs. 
   Nevertheless, we are not excluding a priori that the compact object could be a black hole.
   
   With the available photometric data, the spectral type of the optical counterpart is poorly constrained.
   We showed in Sect. \ref{Sect. Optical counterpart} that, among the other possibilities, \src\ could belong to the class of symbiotic stars,
   with the donor star being a peculiar M-type sub-subgiant star.
   Although symbiotic stars show X-ray flares, these are longer than in \src\ (for example, for GX~1$+$4 $\Delta L_{\rm x}\approx$~days,
   and for 3A~1954$+$319\footnote{whose donor star is an M supergiant, \citep[][ and references therein]{Bozzo22}.}  $\Delta t\approx 10^4$~s), and the dynamic range is smaller
   ($\Delta L_{\rm x}\lesssim 60$;  \citealt{Corbet08, Bozzo22}).

\begin{figure}[ht!]
   \centering
   \includegraphics[width=\columnwidth]{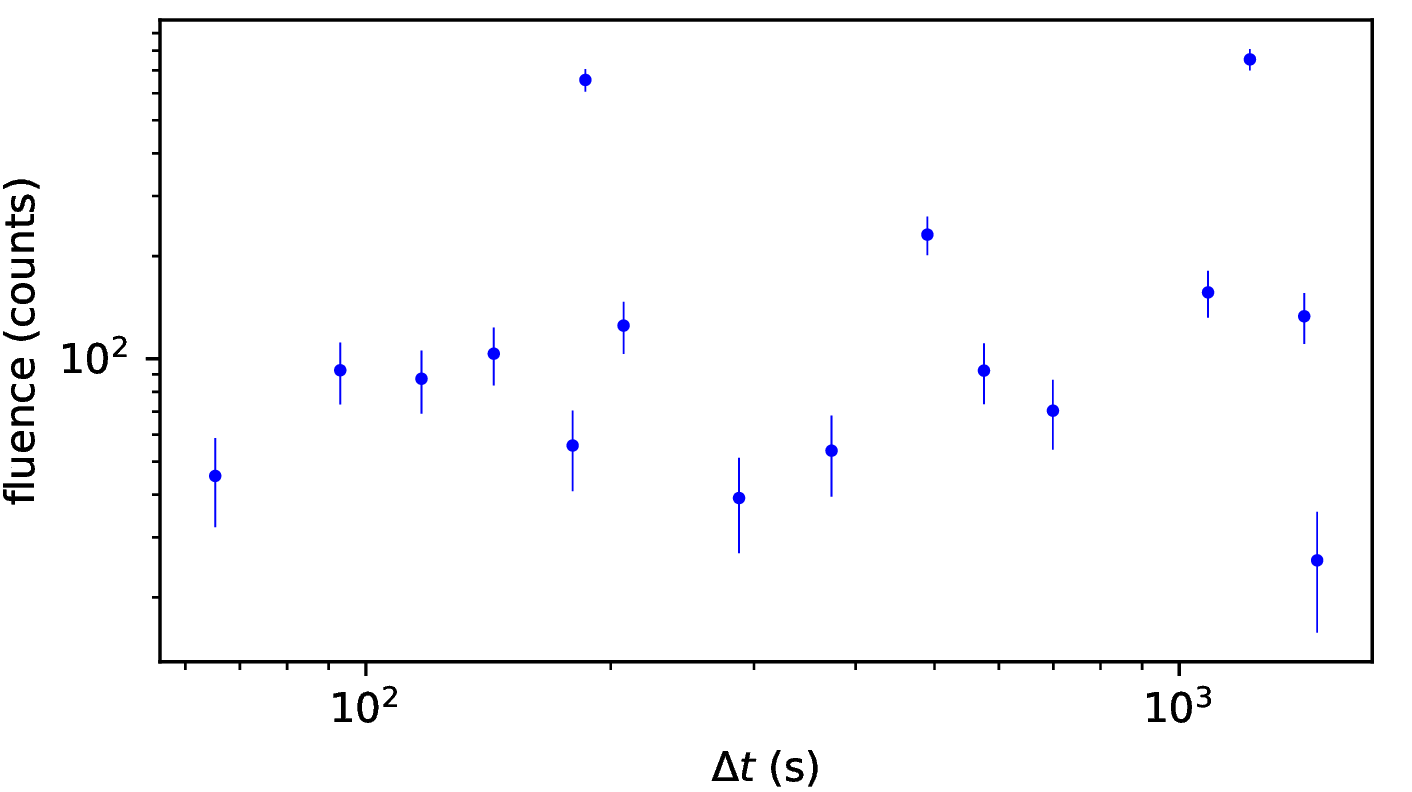}
   \caption{\nustar\ burst fluence (3$-$60~keV) as function of the time interval following the burst, before the next one.}
   \label{nustar fluence}
   \end{figure}

   Our study (Sect. \ref{Sect. Optical counterpart}) and the previous classification of the optical counterpart
   \citep{Greiss11, Kaur11}
   suggest that \src\ might be a LMXB, with M, F, or G type star, from main sequence to giant.
   Among LMXBs, some accreting millisecond pulsars (AMXP; see, e.g., \citealt{Papitto20})
   show flares with the same durations of those displayed by \src, but their dynamic range
   ($\Delta L_{\rm x} \approx 10-50$) is lower than that observed in \src\ (see, for example the case
   of the ``hiccup'' accretion in IGR J18245$-$2452, \citealt{Ferrigno14}).
   Other X-ray binaries showing similar flares are the LMXBs ``Bursting Pulsar'' GRO~J1744$-$28,
   the ``Rapid Burster'' MXB~1730$-$335, and the HMXB SMC~X$-$1 \citep[see, e.g., ][ and references therein]{Bagnoli15, Court18, Rai18}.
   They exhibit \emph{type-II bursts}, which typically last for a few tens of seconds (there are, however, some exceptions; see references above). Generally they are single peaked (although, in some cases, they show more complex structures) and there is no evidence of significant spectral variability between the persistent and the flaring states.
   At odds with \src, these three sources show type-II bursts only during bright luminosity states
   where the persistent X-ray luminosity is $L_{\rm x}\approx 10^{37}-10^{38}$~erg~s$^{-1}$
   and the dynamic range of the flares is $\Delta L_{\rm x} \approx 10-40$ \citep[e.g.][]{Bagnoli15, Giles96, Sazonov97}.
   Different models have been proposed to explain the type-II bursts. Among them, those that can also explain
   why the other NS LMXBs do not show these bursts belong to the family of the trapped disc models
   \citep{Spruit93, Dangelo10, Dangelo12, vandenEijnden17}. In this model, for some specific values of the magnetic field strength
   and mass capture rate ($B\approx 10^{10}-10^{11}$~G, $\dot{M}_{\rm c}$ roughly $\sim 10$\%$-45$\% of the Eddington rate), 
   the interaction between the inner region of the accretion disc and
   the magnetospheric boundary of the pulsar may interrupt the continuous flow and produce a cycle of accretion events (bursts).
   The fluence of the type-II bursts in the Rapid Burster is proportional to the waiting time to the following burst
   \citep{Lewin76}. This is not observed in the Bursting Pulsar \cite{Kouveliotou96} and, to the best of our knowledge, in SMC~X-1.
   In Fig. \ref{nustar fluence} we checked if this is also the case for \src, but we did not find any significant correlation between the fluence and the waiting time.
   For this calculations, we considered only flares and waiting times not interrupted by the regular gaps that characterise the \nustar\ observation, where other flares may have occurred.

  \src\  shows some striking  similarities with the LMXBs Swift~J1858.6$-$0814, V404~Cygni, and V4641~Sgr.
These sources show flares spanning few orders of magnitude ($\sim 10^2-10^3$) on timescales of $\sim 10-100$~s \citep[see, e.g., ][]{Ludlam18,Hare20,Wijnands00,Rodriguez15}. V404~Cygni and V4641~Sgr host BHs, while the compact object in Swift~J1858.6$-$0814
was identified to be a NS, thanks to the detection of some type I X-ray bursts \citep{Buisson20}.
Despite these similarities, these sources show the flaring activity on top of longer (weeks to months) X-ray outbursts,
and the flares show a remarkable spectral variability in the X-ray spectrum, at odds with \src.

   The fast and strong variability of \src\ is also reminiscent of that observed by \xmm\ from the Be/XRBs A0538$-$66 in 2018. 
   This binary system, located in the Large Magellanic Cloud, hosts a $\sim 69$~ms pulsar orbiting around a Be star
   in $P_{\rm orb}\approx 16.6$ days, in a highly eccentric orbit ($e\approx 0.72$; \citealt{Ducci22, Rajoelimanana17}).
   During the \xmm\ observations of 2018, A0538$-$66 showed X-ray flares with durations between 5 and 50 seconds, and peak X-ray luminosity up to $\sim 4\times 10^{38}$~erg~s$^{-1}$ (0.2$-$10~keV).
   Between the flares, the luminosity was $\sim 2\times 10^{35}$~erg~s$^{-1}$.
   Therefore, the dynamic range and flare durations of \src\ and A0538$-$66 are consistent,
   while the luminosities of \src\ are $\sim 6\times 10^2$ times lower than in A0538$-$66, assuming $d=2$~kpc.
   Another similarity between these two sources is that the flares of both show a single-peak or multi-peak structure.
   To explain the X-ray variability of A0538$-$66 observed by \xmm\ in 2018, different plausible mechanism were considered. 
   In particular, a gating mechanism in a regime of spherically symmetric inflow was explored in detail. In the proposed scenario,
   the observed X-ray variability is linked to fast transitions between accretion and supersonic propeller regimes \citep{Ducci19}.
   These accretion regimes and the transitions between them occur under particular conditions,
   which are mainly driven by the spin period of the pulsar, its magnetic field strength, and the rate of the mass
   it gravitationally captures \citep{Davies81, Bozzo08}.
   In the introduction, we reported about an optical brightening of about 1 magnitude in \src\ \citep{Kaur11}.
   Notably, A0538$-$66 also shows fast and bright optical flares ($\Delta t \lesssim 1$~d, $\Delta m_{\rm v}\approx 0.5-1$; rarely up to $\Delta m_{\rm v}\approx 2.2$; \citealt{Ducci16} and references therein).
   They were explained with reprocessing of the X-ray photons from the accreting pulsar
   in a cloud surrounding the binary system, and it was also pointed out that it was possible 
   to have a non-negligible contribution to the optical brightening caused by the heating of the surface of the 
   companion star irradiated by the X-ray pulsar \citep{Ducci19b}.

   \section{Conclusions}

We have reported the results of quasi-simultaneous \nustar\ and \xmm\
observations of \src. These show an astonishing variability, whose properties
are so extreme to be difficult to explain by known accretion mechanisms
applied to other accreting binary systems. We have shown that the variability
of \src\ shows some similarities (and also substantial differences)
with a few LXMBs, in particular Swift~J1858.6$-$0814, and a remarkable
similarity with A0538$-$66, which, however, is an HMXB.
The uniqueness of the X-ray flashes displayed by \src\ requires
further observations in the X-ray, optical, and infrared bands,
to unveil the nature of the stars of the system and to determine 
the mechanism that causes its X-ray variability.
In particular, future spectroscopic observations in optical near-infrared
are fundamental to constrain the properties of the donor star.

   \begin{acknowledgements}
   We thank the anonymous referee for constructive comments that helped to improve the paper.
   LD acknowledges Dr. Marilena Caramazza and Dr. Dante Minniti for their helpful suggestions.
    PR and EB acknowledge financial contribution from the agreement ASI-INAF I/037/12/0. 
    This research has made use of the NuSTAR Data Analysis Software (NuSTARDAS) jointly developed by the ASI Space Science
    Data Center (SSDC, Italy) and the California Institute of Technology (Caltech, USA). 
    Based on observations obtained with XMM-Newton, an ESA science mission with instruments and contributions directly funded by ESA Member States and NASA.
    Part of the data analysis was performed using {\tt stingray}, an open source spectral-timing Python software package
    for astrophysical data analysis \citep{Bachetti22, Huppenkothen19a, Huppenkothen19b}.
    This work made use of Astropy: \url{http://www.astropy.org} a community-developed core Python package
    and an ecosystem of tools and resources for astronomy \citep{astropy:2013, astropy:2018, astropy:2022}. 
    It also made use of {\tt dustmaps}, a unified interface for several 2D and 3D maps of interstellar dust reddening and extinction \citep{Green18}.
    This research has made use of the VizieR catalogue access tool, CDS, Strasbourg, France.
   \end{acknowledgements}

\bibliographystyle{aa} 
\bibliography{ld_j17407}

\begin{thebibliography}{96}
\expandafter\ifx\csname natexlab\endcsname\relax\def\natexlab#1{#1}\fi

\bibitem[{{Anders} \& {Grevesse}(1989)}]{Anders89}
{Anders}, E. \& {Grevesse}, N. 1989, \gca, 53, 197

\bibitem[{{Anitra} {et~al.}(2021){Anitra}, {Di Salvo}, {Iaria}, {Burderi},
  {Gambino}, {Mazzola}, {Marino}, {Sanna}, \& {Riggio}}]{4u1}
{Anitra}, A., {Di Salvo}, T., {Iaria}, R., {et~al.} 2021, \aap, 654, A160

\bibitem[{{Arnaud}(1996)}]{Arnaud96}
{Arnaud}, K.~A. 1996, in Astronomical Society of the Pacific Conference Series,
  Vol. 101, Astronomical Data Analysis Software and Systems V, ed. G.~H.
  {Jacoby} \& J.~{Barnes}, 17

\bibitem[{{Astropy Collaboration} {et~al.}(2022){Astropy Collaboration},
  {Price-Whelan}, {Lim}, {Earl}, {Starkman}, {Bradley}, {Shupe}, {Patil},
  {Corrales}, {Brasseur}, {N{"o}the}, {Donath}, {Tollerud}, {Morris},
  {Ginsburg}, {Vaher}, {Weaver}, {Tocknell}, {Jamieson}, {van Kerkwijk},
  {Robitaille}, {Merry}, {Bachetti}, {G{"u}nther}, {Aldcroft},
  {Alvarado-Montes}, {Archibald}, {B{'o}di}, {Bapat}, {Barentsen}, {Baz{'a}n},
  {Biswas}, {Boquien}, {Burke}, {Cara}, {Cara}, {Conroy}, {Conseil}, {Craig},
  {Cross}, {Cruz}, {D'Eugenio}, {Dencheva}, {Devillepoix}, {Dietrich},
  {Eigenbrot}, {Erben}, {Ferreira}, {Foreman-Mackey}, {Fox}, {Freij}, {Garg},
  {Geda}, {Glattly}, {Gondhalekar}, {Gordon}, {Grant}, {Greenfield}, {Groener},
  {Guest}, {Gurovich}, {Handberg}, {Hart}, {Hatfield-Dodds}, {Homeier},
  {Hosseinzadeh}, {Jenness}, {Jones}, {Joseph}, {Kalmbach}, {Karamehmetoglu},
  {Ka{l}uszy{'n}ski}, {Kelley}, {Kern}, {Kerzendorf}, {Koch}, {Kulumani},
  {Lee}, {Ly}, {Ma}, {MacBride}, {Maljaars}, {Muna}, {Murphy}, {Norman},
  {O'Steen}, {Oman}, {Pacifici}, {Pascual}, {Pascual-Granado}, {Patil},
  {Perren}, {Pickering}, {Rastogi}, {Roulston}, {Ryan}, {Rykoff}, {Sabater},
  {Sakurikar}, {Salgado}, {Sanghi}, {Saunders}, {Savchenko}, {Schwardt},
  {Seifert-Eckert}, {Shih}, {Jain}, {Shukla}, {Sick}, {Simpson},
  {Singanamalla}, {Singer}, {Singhal}, {Sinha}, {Sip{H{o}}cz}, {Spitler},
  {Stansby}, {Streicher}, {{{S}}umak}, {Swinbank}, {Taranu}, {Tewary},
  {Tremblay}, {Val-Borro}, {Van Kooten}, {Vasovi{'c}}, {Verma}, {de Miranda
  Cardoso}, {Williams}, {Wilson}, {Winkel}, {Wood-Vasey}, {Xue}, {Yoachim},
  {Zhang}, {Zonca}, \& {Astropy Project Contributors}}]{astropy:2022}
{Astropy Collaboration}, {Price-Whelan}, A.~M., {Lim}, P.~L., {et~al.} 2022,
  apj, 935, 167

\bibitem[{{Astropy Collaboration} {et~al.}(2018){Astropy Collaboration},
  {Price-Whelan}, {Sip{\H{o}}cz}, {G{\"u}nther}, {Lim}, {Crawford}, {Conseil},
  {Shupe}, {Craig}, {Dencheva}, {Ginsburg}, {Vand erPlas}, {Bradley},
  {P{\'e}rez-Su{\'a}rez}, {de Val-Borro}, {Aldcroft}, {Cruz}, {Robitaille},
  {Tollerud}, {Ardelean}, {Babej}, {Bach}, {Bachetti}, {Bakanov}, {Bamford},
  {Barentsen}, {Barmby}, {Baumbach}, {Berry}, {Biscani}, {Boquien}, {Bostroem},
  {Bouma}, {Brammer}, {Bray}, {Breytenbach}, {Buddelmeijer}, {Burke},
  {Calderone}, {Cano Rodr{\'\i}guez}, {Cara}, {Cardoso}, {Cheedella}, {Copin},
  {Corrales}, {Crichton}, {D'Avella}, {Deil}, {Depagne}, {Dietrich}, {Donath},
  {Droettboom}, {Earl}, {Erben}, {Fabbro}, {Ferreira}, {Finethy}, {Fox},
  {Garrison}, {Gibbons}, {Goldstein}, {Gommers}, {Greco}, {Greenfield},
  {Groener}, {Grollier}, {Hagen}, {Hirst}, {Homeier}, {Horton}, {Hosseinzadeh},
  {Hu}, {Hunkeler}, {Ivezi{\'c}}, {Jain}, {Jenness}, {Kanarek}, {Kendrew},
  {Kern}, {Kerzendorf}, {Khvalko}, {King}, {Kirkby}, {Kulkarni}, {Kumar},
  {Lee}, {Lenz}, {Littlefair}, {Ma}, {Macleod}, {Mastropietro}, {McCully},
  {Montagnac}, {Morris}, {Mueller}, {Mumford}, {Muna}, {Murphy}, {Nelson},
  {Nguyen}, {Ninan}, {N{\"o}the}, {Ogaz}, {Oh}, {Parejko}, {Parley}, {Pascual},
  {Patil}, {Patil}, {Plunkett}, {Prochaska}, {Rastogi}, {Reddy Janga},
  {Sabater}, {Sakurikar}, {Seifert}, {Sherbert}, {Sherwood-Taylor}, {Shih},
  {Sick}, {Silbiger}, {Singanamalla}, {Singer}, {Sladen}, {Sooley},
  {Sornarajah}, {Streicher}, {Teuben}, {Thomas}, {Tremblay}, {Turner},
  {Terr{\'o}n}, {van Kerkwijk}, {de la Vega}, {Watkins}, {Weaver}, {Whitmore},
  {Woillez}, {Zabalza}, \& {Astropy Contributors}}]{astropy:2018}
{Astropy Collaboration}, {Price-Whelan}, A.~M., {Sip{\H{o}}cz}, B.~M., {et~al.}
  2018, \aj, 156, 123

\bibitem[{{Astropy Collaboration} {et~al.}(2013){Astropy Collaboration},
  {Robitaille}, {Tollerud}, {Greenfield}, {Droettboom}, {Bray}, {Aldcroft},
  {Davis}, {Ginsburg}, {Price-Whelan}, {Kerzendorf}, {Conley}, {Crighton},
  {Barbary}, {Muna}, {Ferguson}, {Grollier}, {Parikh}, {Nair}, {Unther},
  {Deil}, {Woillez}, {Conseil}, {Kramer}, {Turner}, {Singer}, {Fox}, {Weaver},
  {Zabalza}, {Edwards}, {Azalee Bostroem}, {Burke}, {Casey}, {Crawford},
  {Dencheva}, {Ely}, {Jenness}, {Labrie}, {Lim}, {Pierfederici}, {Pontzen},
  {Ptak}, {Refsdal}, {Servillat}, \& {Streicher}}]{astropy:2013}
{Astropy Collaboration}, {Robitaille}, T.~P., {Tollerud}, E.~J., {et~al.} 2013,
  \aap, 558, A33

\bibitem[{Bachetti {et~al.}(2022)Bachetti, Huppenkothen, Khan, Mishra, Sharma,
  Stevens, Swinbank, Desai, Rashid, Ribeiro, Tripathi, Sipőcz, Vats, tappina,
  omargamal8, Davis, Rasquinha, Balm, Mumford, Campana, parkma99, Garg, Tandon,
  Hota, Nick, Raj, Mishra, Smith, Mahlke, \& Sachidanand}]{Bachetti22}
Bachetti, M., Huppenkothen, D., Khan, U., {et~al.} 2022,
  StingraySoftware/stingray: v1.1

\bibitem[{{Bagnoli} {et~al.}(2015){Bagnoli}, {in't Zand}, {D'Angelo}, \&
  {Galloway}}]{Bagnoli15}
{Bagnoli}, T., {in't Zand}, J.~J.~M., {D'Angelo}, C.~R., \& {Galloway}, D.~K.
  2015, \mnras, 449, 268

\bibitem[{{Bahramian} \& {Degenaar}(2022)}]{Bahramian22}
{Bahramian}, A. \& {Degenaar}, N. 2022, arXiv e-prints, arXiv:2206.10053

\bibitem[{{Bozzo} {et~al.}(2018){Bozzo}, {Bahramian}, {Ferrigno}, {Sanna},
  {Strader}, {Lewis}, {Russell}, {di Salvo}, {Burderi}, {Riggio}, {Papitto},
  {Gandhi}, \& {Romano}}]{symb1}
{Bozzo}, E., {Bahramian}, A., {Ferrigno}, C., {et~al.} 2018, \aap, 613, A22

\bibitem[{{Bozzo} {et~al.}(2008){Bozzo}, {Falanga}, \& {Stella}}]{Bozzo08}
{Bozzo}, E., {Falanga}, M., \& {Stella}, L. 2008, \apj, 683, 1031

\bibitem[{{Bozzo} {et~al.}(2022{\natexlab{a}}){Bozzo}, {Ferrigno}, {Oskinova},
  \& {Ducci}}]{Bozzo22}
{Bozzo}, E., {Ferrigno}, C., {Oskinova}, L., \& {Ducci}, L. 2022{\natexlab{a}},
  \mnras, 510, 4645

\bibitem[{{Bozzo} {et~al.}(2022{\natexlab{b}}){Bozzo}, {Romano}, {Ferrigno}, \&
  {Oskinova}}]{symb2}
{Bozzo}, E., {Romano}, P., {Ferrigno}, C., \& {Oskinova}, L.
  2022{\natexlab{b}}, \mnras, 513, 42

\bibitem[{{Brazier}(1994)}]{Brazier94}
{Brazier}, K.~T.~S. 1994, \mnras, 268, 709

\bibitem[{{Buccheri} {et~al.}(1983){Buccheri}, {Bennett}, {Bignami}, {Bloemen},
  {Boriakoff}, {Caraveo}, {Hermsen}, {Kanbach}, {Manchester}, {Masnou},
  {Mayer-Hasselwander}, {{\"O}zel}, {Paul}, {Sacco}, {Scarsi}, \&
  {Strong}}]{Buccheri83}
{Buccheri}, R., {Bennett}, K., {Bignami}, G.~F., {et~al.} 1983, \aap, 128, 245

\bibitem[{{Buisson} {et~al.}(2020){Buisson}, {Altamirano}, {Bult}, {Mancuso},
  {G{\"u}ver}, {Jaisawal}, {Hare}, {Albayati}, {Arzoumanian}, {Castro Segura},
  {Chakrabarty}, {Gandhi}, {Guillot}, {Homan}, {Gendreau}, {Jiang},
  {Malacaria}, {Miller}, {{\"O}zbey Arabac{\i}}, {Remillard}, {Strohmayer},
  {Tombesi}, {Tomsick}, {Vincentelli}, \& {Walton}}]{Buisson20}
{Buisson}, D.~J.~K., {Altamirano}, D., {Bult}, P., {et~al.} 2020, \mnras, 499,
  793

\bibitem[{{Campana}(2009)}]{Campana09}
{Campana}, S. 2009, \apj, 699, 1144

\bibitem[{{Cardelli} {et~al.}(1989){Cardelli}, {Clayton}, \&
  {Mathis}}]{Cardelli89}
{Cardelli}, J.~A., {Clayton}, G.~C., \& {Mathis}, J.~S. 1989, \apj, 345, 245

\bibitem[{{Cash}(1979)}]{Cash79}
{Cash}, W. 1979, \apj, 228, 939

\bibitem[{{Chambers} {et~al.}(2016){Chambers}, {Magnier}, {Metcalfe},
  {Flewelling}, {Huber}, {Waters}, {Denneau}, {Draper}, {Farrow}, {Finkbeiner},
  {Holmberg}, {Koppenhoefer}, {Price}, {Rest}, {Saglia}, {Schlafly}, {Smartt},
  {Sweeney}, {Wainscoat}, {Burgett}, {Chastel}, {Grav}, {Heasley}, {Hodapp},
  {Jedicke}, {Kaiser}, {Kudritzki}, {Luppino}, {Lupton}, {Monet}, {Morgan},
  {Onaka}, {Shiao}, {Stubbs}, {Tonry}, {White}, {Ba{\~n}ados}, {Bell},
  {Bender}, {Bernard}, {Boegner}, {Boffi}, {Botticella}, {Calamida},
  {Casertano}, {Chen}, {Chen}, {Cole}, {Deacon}, {Frenk}, {Fitzsimmons},
  {Gezari}, {Gibbs}, {Goessl}, {Goggia}, {Gourgue}, {Goldman}, {Grant},
  {Grebel}, {Hambly}, {Hasinger}, {Heavens}, {Heckman}, {Henderson}, {Henning},
  {Holman}, {Hopp}, {Ip}, {Isani}, {Jackson}, {Keyes}, {Koekemoer}, {Kotak},
  {Le}, {Liska}, {Long}, {Lucey}, {Liu}, {Martin}, {Masci}, {McLean}, {Mindel},
  {Misra}, {Morganson}, {Murphy}, {Obaika}, {Narayan}, {Nieto-Santisteban},
  {Norberg}, {Peacock}, {Pier}, {Postman}, {Primak}, {Rae}, {Rai}, {Riess},
  {Riffeser}, {Rix}, {R{\"o}ser}, {Russel}, {Rutz}, {Schilbach}, {Schultz},
  {Scolnic}, {Strolger}, {Szalay}, {Seitz}, {Small}, {Smith}, {Soderblom},
  {Taylor}, {Thomson}, {Taylor}, {Thakar}, {Thiel}, {Thilker}, {Unger},
  {Urata}, {Valenti}, {Wagner}, {Walder}, {Walter}, {Watters}, {Werner},
  {Wood-Vasey}, \& {Wyse}}]{Chambers16}
{Chambers}, K.~C., {Magnier}, E.~A., {Metcalfe}, N., {et~al.} 2016, arXiv
  e-prints, arXiv:1612.05560

\bibitem[{{Corbet} {et~al.}(2008){Corbet}, {Sokoloski}, {Mukai}, {Markwardt},
  \& {Tueller}}]{Corbet08}
{Corbet}, R.~H.~D., {Sokoloski}, J.~L., {Mukai}, K., {Markwardt}, C.~B., \&
  {Tueller}, J. 2008, \apj, 675, 1424

\bibitem[{{Cornelisse} {et~al.}(2004){Cornelisse}, {in't Zand}, {Kuulkers},
  {Heise}, {Verbunt}, {Cocchi}, {Bazzano}, {Natalucci}, \&
  {Ubertini}}]{Cornelisse04}
{Cornelisse}, R., {in't Zand}, J.~J.~M., {Kuulkers}, E., {et~al.} 2004, Nuclear
  Physics B Proceedings Supplements, 132, 518

\bibitem[{{Court} {et~al.}(2018){Court}, {Altamirano}, {Albayati}, {Sanna},
  {Belloni}, {Overton}, {Degenaar}, {Wijnands}, {Yamaoka}, {Hill}, \&
  {Knigge}}]{Court18}
{Court}, J.~M.~C., {Altamirano}, D., {Albayati}, A.~C., {et~al.} 2018, \mnras,
  481, 2273

\bibitem[{{D'A{\`\i}} {et~al.}(2015){D'A{\`\i}}, {Di Salvo}, {Iaria},
  {Garc{\'\i}a}, {Sanna}, {Pintore}, {Riggio}, {Burderi}, {Bozzo}, {Dauser},
  {Matranga}, {Galiano}, \& {Robba}}]{gro}
{D'A{\`\i}}, A., {Di Salvo}, T., {Iaria}, R., {et~al.} 2015, \mnras, 449, 4288

\bibitem[{{D'Angelo} \& {Spruit}(2010)}]{Dangelo10}
{D'Angelo}, C.~R. \& {Spruit}, H.~C. 2010, \mnras, 406, 1208

\bibitem[{{D'Angelo} \& {Spruit}(2012)}]{Dangelo12}
{D'Angelo}, C.~R. \& {Spruit}, H.~C. 2012, \mnras, 420, 416

\bibitem[{{Davies} \& {Pringle}(1981)}]{Davies81}
{Davies}, R.~E. \& {Pringle}, J.~E. 1981, \mnras, 196, 209

\bibitem[{{de Jager} \& {Nieuwenhuijzen}(1987)}]{deJager87}
{de Jager}, C. \& {Nieuwenhuijzen}, H. 1987, \aap, 177, 217

\bibitem[{{Del Santo} {et~al.}(2010){Del Santo}, {Sidoli}, {Romano}, {Bazzano},
  {Wijnands}, {Degenaar}, \& {Mereghetti}}]{DelSanto10}
{Del Santo}, M., {Sidoli}, L., {Romano}, P., {et~al.} 2010, \mnras, 403, L89

\bibitem[{{Done} {et~al.}(2007){Done}, {Gierli{\'n}ski}, \& {Kubota}}]{Done07}
{Done}, C., {Gierli{\'n}ski}, M., \& {Kubota}, A. 2007, \aapr, 15, 1

\bibitem[{{Ducci} {et~al.}(2016){Ducci}, {Covino}, {Doroshenko}, {Mereghetti},
  {Santangelo}, \& {Sasaki}}]{Ducci16}
{Ducci}, L., {Covino}, S., {Doroshenko}, V., {et~al.} 2016, \aap, 595, A103

\bibitem[{{Ducci} {et~al.}(2019{\natexlab{a}}){Ducci}, {Mereghetti},
  {Hryniewicz}, {Santangelo}, \& {Romano}}]{Ducci19b}
{Ducci}, L., {Mereghetti}, S., {Hryniewicz}, K., {Santangelo}, A., \& {Romano},
  P. 2019{\natexlab{a}}, \aap, 624, A9

\bibitem[{{Ducci} {et~al.}(2019{\natexlab{b}}){Ducci}, {Mereghetti}, \&
  {Santangelo}}]{Ducci19}
{Ducci}, L., {Mereghetti}, S., \& {Santangelo}, A. 2019{\natexlab{b}}, \apjl,
  881, L17

\bibitem[{{Ducci} {et~al.}(2022){Ducci}, {Mereghetti}, {Santangelo}, {Ji},
  {Carpano}, {Covino}, {Doroshenko}, {Haberl}, {Maitra}, {Kreykenbohm}, \&
  {Udalski}}]{Ducci22}
{Ducci}, L., {Mereghetti}, S., {Santangelo}, A., {et~al.} 2022, \aap, 661, A22

\bibitem[{{Ducci} {et~al.}(2010){Ducci}, {Sidoli}, \& {Paizis}}]{Ducci10}
{Ducci}, L., {Sidoli}, L., \& {Paizis}, A. 2010, \mnras, 408, 1540

\bibitem[{{Ferrigno} {et~al.}(2014){Ferrigno}, {Bozzo}, {Papitto}, {Rea},
  {Pavan}, {Campana}, {Wieringa}, {Filipovi{\'c}}, {Falanga}, \&
  {Stella}}]{Ferrigno14}
{Ferrigno}, C., {Bozzo}, E., {Papitto}, A., {et~al.} 2014, \aap, 567, A77

\bibitem[{{Foight} {et~al.}(2016){Foight}, {G{\"u}ver}, {{\"O}zel}, \&
  {Slane}}]{Foight16}
{Foight}, D.~R., {G{\"u}ver}, T., {{\"O}zel}, F., \& {Slane}, P.~O. 2016, \apj,
  826, 66

\bibitem[{{Giles} {et~al.}(1996){Giles}, {Swank}, {Jahoda}, {Zhang},
  {Strohmayer}, {Stark}, \& {Morgan}}]{Giles96}
{Giles}, A.~B., {Swank}, J.~H., {Jahoda}, K., {et~al.} 1996, \apjl, 469, L25

\bibitem[{{Gonzalez} {et~al.}(2011){Gonzalez}, {Rejkuba}, {Zoccali}, {Valenti},
  \& {Minniti}}]{Gonzalez11}
{Gonzalez}, O.~A., {Rejkuba}, M., {Zoccali}, M., {Valenti}, E., \& {Minniti},
  D. 2011, \aap, 534, A3

\bibitem[{{G\"otz} {et~al.}(2004){G\"otz}, {Mereghetti}, {Mowlavi}, \&
  {Soldan}}]{Gotz04}
{G\"otz}, D., {Mereghetti}, S., {Mowlavi}, N., \& {Soldan}, J. 2004, GRB
  Coordinates Network, 2793, 1

\bibitem[{{Grebenev} \& {Sunyaev}(2007)}]{Grebenev07}
{Grebenev}, S.~A. \& {Sunyaev}, R.~A. 2007, Astronomy Letters, 33, 149

\bibitem[{{Green}(2018)}]{Green18}
{Green}, G. 2018, The Journal of Open Source Software, 3, 695

\bibitem[{{Green} {et~al.}(2019){Green}, {Schlafly}, {Zucker}, {Speagle}, \&
  {Finkbeiner}}]{Green19}
{Green}, G.~M., {Schlafly}, E., {Zucker}, C., {Speagle}, J.~S., \&
  {Finkbeiner}, D. 2019, \apj, 887, 93

\bibitem[{{Greiss} {et~al.}(2011){Greiss}, {Steeghs}, {Maccarone}, {Hynes},
  {Britt}, {Jonker}, {Torres}, {Masetti}, {Rojas}, {Heinke}, {Kaur}, {Bird},
  {Gbs Consortium}, \& {Vvv Consortium}}]{Greiss11}
{Greiss}, S., {Steeghs}, D., {Maccarone}, T., {et~al.} 2011, The Astronomer's
  Telegram, 3688, 1

\bibitem[{{Hare} {et~al.}(2020){Hare}, {Tomsick}, {Buisson}, {Clavel},
  {Gandhi}, {Garc{\'\i}a}, {Grefenstette}, {Walton}, \& {Xu}}]{Hare20}
{Hare}, J., {Tomsick}, J.~A., {Buisson}, D. J.~K., {et~al.} 2020, \apj, 890, 57

\bibitem[{{Harrison} {et~al.}(2013){Harrison}, {Craig}, {Christensen},
  {Hailey}, {Zhang}, {Boggs}, {Stern}, {Cook}, {Forster}, {Giommi},
  {Grefenstette}, {Kim}, {Kitaguchi}, {Koglin}, {Madsen}, {Mao}, {Miyasaka},
  {Mori}, {Perri}, {Pivovaroff}, {Puccetti}, {Rana}, {Westergaard}, {Willis},
  {Zoglauer}, {An}, {Bachetti}, {Barri{\`e}re}, {Bellm}, {Bhalerao},
  {Brejnholt}, {Fuerst}, {Liebe}, {Markwardt}, {Nynka}, {Vogel}, {Walton},
  {Wik}, {Alexander}, {Cominsky}, {Hornschemeier}, {Hornstrup}, {Kaspi},
  {Madejski}, {Matt}, {Molendi}, {Smith}, {Tomsick}, {Ajello}, {Ballantyne},
  {Balokovi{\'c}}, {Barret}, {Bauer}, {Blandford}, {Brandt}, {Brenneman},
  {Chiang}, {Chakrabarty}, {Chenevez}, {Comastri}, {Dufour}, {Elvis}, {Fabian},
  {Farrah}, {Fryer}, {Gotthelf}, {Grindlay}, {Helfand}, {Krivonos}, {Meier},
  {Miller}, {Natalucci}, {Ogle}, {Ofek}, {Ptak}, {Reynolds}, {Rigby},
  {Tagliaferri}, {Thorsett}, {Treister}, \& {Urry}}]{Harrison13}
{Harrison}, F.~A., {Craig}, W.~W., {Christensen}, F.~E., {et~al.} 2013, \apj,
  770, 103

\bibitem[{{Heinke} {et~al.}(2009){Heinke}, {Tomsick}, {Yusef-Zadeh}, \&
  {Grindlay}}]{Heinke09}
{Heinke}, C.~O., {Tomsick}, J.~A., {Yusef-Zadeh}, F., \& {Grindlay}, J.~E.
  2009, \apj, 701, 1627

\bibitem[{{Herpich} {et~al.}(2021){Herpich}, {Ferreira Lopes}, {Saito},
  {Minniti}, {Ederoclite}, {Ferreira}, \& {Catelan}}]{Herpich21}
{Herpich}, F.~R., {Ferreira Lopes}, C.~E., {Saito}, R.~K., {et~al.} 2021, \aap,
  647, A169

\bibitem[{{Huppenkothen} {et~al.}(2019){Huppenkothen}, {Bachetti}, {Stevens},
  {Migliari}, {Balm}, {Hammad}, {Khan}, {Mishra}, {Rashid}, {Sharma}, {Martinez
  Ribeiro}, \& {Valles Blanco}}]{Huppenkothen19a}
{Huppenkothen}, D., {Bachetti}, M., {Stevens}, A.~L., {et~al.} 2019, ApJ, 881,
  39

\bibitem[{Huppenkothen {et~al.}(2019)Huppenkothen, Bachetti, Stevens, Migliari,
  Balm, Hammad, Khan, Mishra, Rashid, Sharma, Ribeiro, \&
  Blanco}]{Huppenkothen19b}
Huppenkothen, D., Bachetti, M., Stevens, A.~L., {et~al.} 2019, Journal of Open
  Source Software, 4, 1393

\bibitem[{{in't Zand}(2005)}]{intZand05}
{in't Zand}, J.~J.~M. 2005, \aap, 441, L1

\bibitem[{{Jansen} {et~al.}(2001){Jansen}, {Lumb}, {Altieri}, {Clavel}, {Ehle},
  {Erd}, {Gabriel}, {Guainazzi}, {Gondoin}, {Much}, {Munoz}, {Santos},
  {Schartel}, {Texier}, \& {Vacanti}}]{Jansen01}
{Jansen}, F., {Lumb}, D., {Altieri}, B., {et~al.} 2001, \aap, 365, L1

\bibitem[{{Jonker} {et~al.}(2003){Jonker}, {van der Klis}, \& {Groot}}]{4u2}
{Jonker}, P.~G., {van der Klis}, M., \& {Groot}, P.~J. 2003, \mnras, 339, 663

\bibitem[{{Kaur} {et~al.}(2011){Kaur}, {Heinke}, {Kotulla}, {Eigenbrot},
  {Schechtman-Rook}, \& {Kaplan}}]{Kaur11}
{Kaur}, R., {Heinke}, C., {Kotulla}, R., {et~al.} 2011, The Astronomer's
  Telegram, 3695, 1

\bibitem[{{King} \& {Wijnands}(2006)}]{King06}
{King}, A.~R. \& {Wijnands}, R. 2006, \mnras, 366, L31

\bibitem[{{Kouveliotou} {et~al.}(1996){Kouveliotou}, {van Paradijs}, {Fishman},
  {Briggs}, {Kommers}, {Harmon}, {Meegan}, \& {Lewin}}]{Kouveliotou96}
{Kouveliotou}, C., {van Paradijs}, J., {Fishman}, G.~J., {et~al.} 1996, \nat,
  379, 799

\bibitem[{{Kretschmar} {et~al.}(2019){Kretschmar}, {F{\"u}rst}, {Sidoli},
  {Bozzo}, {Alfonso-Garz{\'o}n}, {Bodaghee}, {Chaty}, {Chernyakova},
  {Ferrigno}, {Manousakis}, {Negueruela}, {Postnov}, {Paizis}, {Reig},
  {Rodes-Roca}, {Tsygankov}, {Bird}, {Bissinger n{\'e} K{\"u}hnel}, {Blay},
  {Caballero}, {Coe}, {Domingo}, {Doroshenko}, {Ducci}, {Falanga}, {Grebenev},
  {Grinberg}, {Hemphill}, {Kreykenbohm}, {Kreykenbohm n{\'e} Fritz}, {Li},
  {Lutovinov}, {Mart{\'\i}nez-N{\'u}{\~n}ez}, {Mas-Hesse}, {Masetti},
  {McBride}, {Neronov}, {Pottschmidt}, {Rodriguez}, {Romano}, {Rothschild},
  {Santangelo}, {Sguera}, {Staubert}, {Tomsick}, {Torrej{\'o}n}, {Torres},
  {Walter}, {Wilms}, {Wilson-Hodge}, \& {Zhang}}]{Kretschmar19}
{Kretschmar}, P., {F{\"u}rst}, F., {Sidoli}, L., {et~al.} 2019, \nar, 86,
  101546

\bibitem[{{Kretschmar} {et~al.}(2004){Kretschmar}, {Mereghetti}, {Hermsen},
  {Ubertini}, {Winkler}, {Brandt}, \& {Diehl}}]{Kretschmar04}
{Kretschmar}, P., {Mereghetti}, S., {Hermsen}, W., {et~al.} 2004, The
  Astronomer's Telegram, 345, 1

\bibitem[{{Lewin} {et~al.}(1976){Lewin}, {Doty}, {Clark}, {Rappaport}, {Bradt},
  {Doxsey}, {Hearn}, {Hoffman}, {Jernigan}, {Li}, {Mayer}, {McClintock},
  {Primini}, \& {Richardson}}]{Lewin76}
{Lewin}, W.~H.~G., {Doty}, J., {Clark}, G.~W., {et~al.} 1976, \apjl, 207, L95

\bibitem[{{Ludlam} {et~al.}(2018){Ludlam}, {Miller}, {Arzoumanian}, {Gendreau},
  {Bult}, {Strohmayer}, {Markwardt}, {Homan}, {Uttley}, {Cackett},
  {Chakrabarty}, {Altamirano}, {Steiner}, {Jaisawal}, {Guillot}, {Wolff},
  {Ray}, \& {Fabian}}]{Ludlam18}
{Ludlam}, R.~M., {Miller}, J.~M., {Arzoumanian}, Z., {et~al.} 2018, The
  Astronomer's Telegram, 12158, 1

\bibitem[{{Madsen} {et~al.}(2022){Madsen}, {Forster}, {Grefenstette},
  {Harrison}, \& {Miyasaka}}]{Madsen22}
{Madsen}, K.~K., {Forster}, K., {Grefenstette}, B., {Harrison}, F.~A., \&
  {Miyasaka}, H. 2022, Journal of Astronomical Telescopes, Instruments, and
  Systems, 8, 034003

\bibitem[{{Magnier} {et~al.}(2020){Magnier}, {Schlafly}, {Finkbeiner}, {Tonry},
  {Goldman}, {R{\"o}ser}, {Schilbach}, {Casertano}, {Chambers}, {Flewelling},
  {Huber}, {Price}, {Sweeney}, {Waters}, {Denneau}, {Draper}, {Hodapp},
  {Jedicke}, {Kaiser}, {Kudritzki}, {Metcalfe}, {Stubbs}, \&
  {Wainscoat}}]{Magnier20}
{Magnier}, E.~A., {Schlafly}, E.~F., {Finkbeiner}, D.~P., {et~al.} 2020, \apjs,
  251, 6

\bibitem[{{Mereminskiy} {et~al.}(2020){Mereminskiy}, {Lutovinov}, {Sunyaev},
  {Grebenev}, {Krivonos}, \& {Molkov}}]{Mereminskiy20}
{Mereminskiy}, I., {Lutovinov}, A., {Sunyaev}, R., {et~al.} 2020, The
  Astronomer's Telegram, 13630, 1

\bibitem[{{Minniti} {et~al.}(2017){Minniti}, {Lucas}, \& {VVV
  Team}}]{Minniti17}
{Minniti}, D., {Lucas}, P., \& {VVV Team}. 2017, VizieR Online Data Catalog,
  II/348

\bibitem[{{O'Donnell}(1994)}]{ODonnell94}
{O'Donnell}, J.~E. 1994, \apj, 422, 158

\bibitem[{{Papitto} {et~al.}(2020){Papitto}, {Falanga}, {Hermsen},
  {Mereghetti}, {Kuiper}, {Poutanen}, {Bozzo}, {Ambrosino}, {Coti Zelati}, {De
  Falco}, {de Martino}, {Di Salvo}, {Esposito}, {Ferrigno}, {Forot},
  {G{\"o}tz}, {Gouiffes}, {Iaria}, {Laurent}, {Li}, {Li}, {Mineo}, {Moran},
  {Neronov}, {Paizis}, {Rea}, {Riggio}, {Sanna}, {Savchenko}, {S{\l}owikowska},
  {Shearer}, {Tiengo}, \& {Torres}}]{Papitto20}
{Papitto}, A., {Falanga}, M., {Hermsen}, W., {et~al.} 2020, \nar, 91, 101544

\bibitem[{{Pecaut} \& {Mamajek}(2013)}]{Pecaut13}
{Pecaut}, M.~J. \& {Mamajek}, E.~E. 2013, \apjs, 208, 9

\bibitem[{{Rai} {et~al.}(2018){Rai}, {Pradhan}, \& {Paul}}]{Rai18}
{Rai}, B., {Pradhan}, P., \& {Paul}, B.~C. 2018, Research in Astronomy and
  Astrophysics, 18, 148

\bibitem[{{Rajoelimanana} {et~al.}(2017){Rajoelimanana}, {Charles}, {Meintjes},
  {Townsend}, {Schurch}, \& {Udalski}}]{Rajoelimanana17}
{Rajoelimanana}, A.~F., {Charles}, P.~A., {Meintjes}, P.~J., {et~al.} 2017,
  \mnras, 464, 4133

\bibitem[{{Rodriguez} {et~al.}(2015){Rodriguez}, {Cadolle Bel},
  {Alfonso-Garz{\'o}n}, {Siegert}, {Zhang}, {Grinberg}, {Savchenko}, {Tomsick},
  {Chenevez}, {Clavel}, {Corbel}, {Diehl}, {Domingo}, {Gouiff{\`e}s},
  {Greiner}, {Krause}, {Laurent}, {Loh}, {Markoff}, {Mas-Hesse},
  {Miller-Jones}, {Russell}, \& {Wilms}}]{Rodriguez15}
{Rodriguez}, J., {Cadolle Bel}, M., {Alfonso-Garz{\'o}n}, J., {et~al.} 2015,
  \aap, 581, L9

\bibitem[{{Romano}(2015)}]{Romano15}
{Romano}, P. 2015, Journal of High Energy Astrophysics, 7, 126

\bibitem[{{Romano} {et~al.}(2011){Romano}, {Barthelmy}, {Krimm}, {Esposito},
  {Mangano}, {Siegel}, {Burrows}, {Evans}, {Farinelli}, {Kennea}, {Palmer},
  {Vercellone}, \& {Gehrels}}]{Romano11}
{Romano}, P., {Barthelmy}, S.~D., {Krimm}, H.~A., {et~al.} 2011, The
  Astronomer's Telegram, 3685, 1

\bibitem[{{Romano} {et~al.}(2016){Romano}, {Bozzo}, {Esposito}, {Sbarufatti},
  {Haberl}, {Ponti}, {D'Avanzo}, {Ducci}, {Segreto}, {Jin}, {Masetti}, {Del
  Santo}, {Campana}, \& {Mangano}}]{Romano16}
{Romano}, P., {Bozzo}, E., {Esposito}, P., {et~al.} 2016, \aap, 593, A96

\bibitem[{{Romano} {et~al.}(2015){Romano}, {Bozzo}, {Mangano}, {Esposito},
  {Israel}, {Tiengo}, {Campana}, {Ducci}, {Ferrigno}, \& {Kennea}}]{Romano15b}
{Romano}, P., {Bozzo}, E., {Mangano}, V., {et~al.} 2015, \aap, 576, L4

\bibitem[{{Romano} {et~al.}(2014){Romano}, {Ducci}, {Mangano}, {Esposito},
  {Bozzo}, \& {Vercellone}}]{Romano14}
{Romano}, P., {Ducci}, L., {Mangano}, V., {et~al.} 2014, \aap, 568, A55

\bibitem[{{Romano} {et~al.}(2023){Romano}, {Evans}, {Bozzo}, {Mangano},
  {Vercellone}, {Guidorzi}, {Ducci}, {Kennea}, {Barthelmy}, {Palmer}, {Krimm},
  \& {Cenko}}]{Romano23}
{Romano}, P., {Evans}, P.~A., {Bozzo}, E., {et~al.} 2023, \aap, 670, A127

\bibitem[{{Sazonov} {et~al.}(1997){Sazonov}, {Sunyaev}, \& {Lund}}]{Sazonov97}
{Sazonov}, S.~Y., {Sunyaev}, R.~A., \& {Lund}, N. 1997, Astronomy Letters, 23,
  286

\bibitem[{{Scargle} {et~al.}(2013){Scargle}, {Norris}, {Jackson}, \&
  {Chiang}}]{Scargle13}
{Scargle}, J.~D., {Norris}, J.~P., {Jackson}, B., \& {Chiang}, J. 2013, \apj,
  764, 167

\bibitem[{{Sguera} {et~al.}(2006){Sguera}, {Bazzano}, {Bird}, {Dean},
  {Ubertini}, {Barlow}, {Bassani}, {Clark}, {Hill}, {Malizia}, {Molina}, \&
  {Stephen}}]{Sguera06}
{Sguera}, V., {Bazzano}, A., {Bird}, A.~J., {et~al.} 2006, \apj, 646, 452

\bibitem[{{Shakura} {et~al.}(2014){Shakura}, {Postnov}, {Sidoli}, \&
  {Paizis}}]{Shakura14}
{Shakura}, N., {Postnov}, K., {Sidoli}, L., \& {Paizis}, A. 2014, \mnras, 442,
  2325

\bibitem[{{Shaw} {et~al.}(2017){Shaw}, {Heinke}, {Degenaar}, {Wijnands},
  {Kaur}, \& {Forestell}}]{Shaw17}
{Shaw}, A.~W., {Heinke}, C.~O., {Degenaar}, N., {et~al.} 2017, \mnras, 471,
  2508

\bibitem[{{Sidoli}(2017)}]{Sidoli17}
{Sidoli}, L. 2017, in XII Multifrequency Behaviour of High Energy Cosmic
  Sources Workshop (MULTIF2017), 52

\bibitem[{{Sidoli} {et~al.}(2001){Sidoli}, {Belloni}, \&
  {Mereghetti}}]{Sidoli01}
{Sidoli}, L., {Belloni}, T., \& {Mereghetti}, S. 2001, \aap, 368, 835

\bibitem[{{Smith} {et~al.}(2018){Smith}, {Lucas}, {Kurtev}, {Smart}, {Minniti},
  {Borissova}, {Jones}, {Zhang}, {Marocco}, {Contreras Pe{\~n}a}, {Gromadzki},
  {Kuhn}, {Drew}, {Pinfield}, \& {Bedin}}]{Smith18}
{Smith}, L.~C., {Lucas}, P.~W., {Kurtev}, R., {et~al.} 2018, \mnras, 474, 1826

\bibitem[{{Spruit} \& {Taam}(1993)}]{Spruit93}
{Spruit}, H.~C. \& {Taam}, R.~E. 1993, \apj, 402, 593

\bibitem[{{Staubert} {et~al.}(2019){Staubert}, {Tr{\"u}mper}, {Kendziorra},
  {Klochkov}, {Postnov}, {Kretschmar}, {Pottschmidt}, {Haberl}, {Rothschild},
  {Santangelo}, {Wilms}, {Kreykenbohm}, \& {F{\"u}rst}}]{Staubert19}
{Staubert}, R., {Tr{\"u}mper}, J., {Kendziorra}, E., {et~al.} 2019, \aap, 622,
  A61

\bibitem[{{Str{\"u}der} {et~al.}(2001){Str{\"u}der}, {Briel}, {Dennerl},
  {Hartmann}, {Kendziorra}, {Meidinger}, {Pfeffermann}, {Reppin}, {Aschenbach},
  {Bornemann}, {Br{\"a}uninger}, {Burkert}, {Elender}, {Freyberg}, {Haberl},
  {Hartner}, {Heuschmann}, {Hippmann}, {Kastelic}, {Kemmer}, {Kettenring},
  {Kink}, {Krause}, {M{\"u}ller}, {Oppitz}, {Pietsch}, {Popp}, {Predehl},
  {Read}, {Stephan}, {St{\"o}tter}, {Tr{\"u}mper}, {Holl}, {Kemmer}, {Soltau},
  {St{\"o}tter}, {Weber}, {Weichert}, {von Zanthier}, {Carathanassis}, {Lutz},
  {Richter}, {Solc}, {B{\"o}ttcher}, {Kuster}, {Staubert}, {Abbey}, {Holland},
  {Turner}, {Balasini}, {Bignami}, {La Palombara}, {Villa}, {Buttler},
  {Gianini}, {Lain{\'e}}, {Lumb}, \& {Dhez}}]{Strueder01}
{Str{\"u}der}, L., {Briel}, U., {Dennerl}, K., {et~al.} 2001, \aap, 365, L18

\bibitem[{{Tomsick} {et~al.}(2008){Tomsick}, {Chaty}, {Rodriguez}, {Walter}, \&
  {Kaaret}}]{Tomsick08}
{Tomsick}, J.~A., {Chaty}, S., {Rodriguez}, J., {Walter}, R., \& {Kaaret}, P.
  2008, \apj, 685, 1143

\bibitem[{{Turner} {et~al.}(2001){Turner}, {Abbey}, {Arnaud}, {Balasini},
  {Barbera}, {Belsole}, {Bennie}, {Bernard}, {Bignami}, {Boer}, {Briel},
  {Butler}, {Cara}, {Chabaud}, {Cole}, {Collura}, {Conte}, {Cros}, {Denby},
  {Dhez}, {Di Coco}, {Dowson}, {Ferrando}, {Ghizzardi}, {Gianotti}, {Goodall},
  {Gretton}, {Griffiths}, {Hainaut}, {Hochedez}, {Holland}, {Jourdain},
  {Kendziorra}, {Lagostina}, {Laine}, {La Palombara}, {Lortholary}, {Lumb},
  {Marty}, {Molendi}, {Pigot}, {Poindron}, {Pounds}, {Reeves}, {Reppin},
  {Rothenflug}, {Salvetat}, {Sauvageot}, {Schmitt}, {Sembay}, {Short},
  {Spragg}, {Stephen}, {Str{\"u}der}, {Tiengo}, {Trifoglio}, {Tr{\"u}mper},
  {Vercellone}, {Vigroux}, {Villa}, {Ward}, {Whitehead}, \& {Zonca}}]{Turner01}
{Turner}, M.~J.~L., {Abbey}, A., {Arnaud}, M., {et~al.} 2001, \aap, 365, L27

\bibitem[{{Vallenari} {et~al.}(2022){Vallenari}, {Brown}, {Prusti}, {de
  Bruijne}, {Arenou}, {Babusiaux}, {Biermann}, {Creevey}, {Ducourant}, {Evans},
  {Eyer}, {Guerra}, {Hutton}, {Jordi}, {Klioner}, {Lammers}, {Lindegren},
  {Luri}, {Mignard}, {Panem}, {Pourbaix}, {Randich}, {Sartoretti}, {Soubiran},
  {Tanga}, {Walton}, {Bailer-Jones}, {Bastian}, {Drimmel}, {Jansen}, {Katz},
  {Lattanzi}, {van Leeuwen}, {Bakker}, {Cacciari}, {Casta{\~n}eda}, {De
  Angeli}, {Fabricius}, {Fouesneau}, {Fr{\'e}mat}, {Galluccio}, {Guerrier},
  {Heiter}, {Masana}, {Messineo}, {Mowlavi}, {Nicolas}, {Nienartowicz},
  {Pailler}, {Panuzzo}, {Riclet}, {Roux}, {Seabroke}, {Sordo{\o}rcit},
  {Th{\'e}venin}, {Gracia-Abril}, {Portell}, {Teyssier}, {Altmann}, {Andrae},
  {Audard}, {Bellas-Velidis}, {Benson}, {Berthier}, {Blomme}, {Burgess},
  {Busonero}, {Busso}, {C{\'a}novas}, {Carry}, {Cellino}, {Cheek},
  {Clementini}, {Damerdji}, {Davidson}, {de Teodoro}, {Nu{\~n}ez Campos},
  {Delchambre}, {Dell'Oro}, {Esquej}, {Fern{\'a}ndez-Hern{\'a}ndez}, {Fraile},
  {Garabato}, {Garc{\'\i}a-Lario}, {Gosset}, {Haigron}, {Halbwachs}, {Hambly},
  {Harrison}, {Hern{\'a}ndez}, {Hestroffer}, {Hodgkin}, {Holl}, {Jan{\ss}en},
  {Jevardat de Fombelle}, {Jordan}, {Krone-Martins}, {Lanzafame},
  {L{\"o}ffler}, {Marchal}, {Marrese}, {Moitinho}, {Muinonen}, {Osborne},
  {Pancino}, {Pauwels}, {Recio-Blanco}, {Reyl{\'e}}, {Riello}, {Rimoldini},
  {Roegiers}, {Rybizki}, {Sarro}, {Siopis}, {Smith}, {Sozzetti}, {Utrilla},
  {van Leeuwen}, {Abbas}, {{\'A}brah{\'a}m}, {Abreu Aramburu}, {Aerts},
  {Aguado}, {Ajaj}, {Aldea-Montero}, {Altavilla}, {{\'A}lvarez}, {Alves},
  {Anders}, {Anderson}, {Anglada Varela}, {Antoja}, {Baines}, {Baker},
  {Balaguer-N{\'u}{\~n}ez}, {Balbinot}, {Balog}, {Barache}, {Barbato},
  {Barros}, {Barstow}, {Bartolom{\'e}}, {Bassilana}, {Bauchet}, {Becciani},
  {Bellazzini}, {Berihuete}, {Bernet}, {Bertone}, {Bianchi}, {Binnenfeld},
  {Blanco-Cuaresma}, {Blazere}, {Boch}, {Bombrun}, {Bossini}, {Bouquillon},
  {Bragaglia}, {Bramante}, {Breedt}, {Bressan}, {Brouillet}, {Brugaletta},
  {Bucciarelli}, {Burlacu}, {Butkevich}, {Buzzi}, {Caffau}, {Cancelliere},
  {Cantat-Gaudin}, {Carballo}, {Carlucci}, {Carnerero}, {Carrasco},
  {Casamiquela}, {Castellani}, {Castro-Ginard}, {Chaoul}, {Charlot}, {Chemin},
  {Chiaramida}, {Chiavassa}, {Chornay}, {Comoretto}, {Contursi}, {Cooper},
  {Cornez}, {Cowell}, {Crifo}, {Cropper}, {Crosta}, {Crowley}, {Dafonte},
  {Dapergolas}, {David}, {David}, {de Laverny}, {De Luise}, {De March}, {De
  Ridder}, {de Souza}, {de Torres}, {del Peloso}, {del Pozo}, {Delbo},
  {Delgado}, {Delisle}, {Demouchy}, {Dharmawardena}, {Di Matteo}, {Diakite},
  {Diener}, {Distefano}, {Dolding}, {Edvardsson}, {Enke}, {Fabre}, {Fabrizio},
  {Faigler}, {Fedorets}, {Fernique}, {Fienga}, {Figueras}, {Fournier},
  {Fouron}, {Fragkoudi}, {Gai}, {Garcia-Gutierrez}, {Garcia-Reinaldos},
  {Garc{\'\i}a-Torres}, {Garofalo}, {Gavel}, {Gavras}, {Gerlach}, {Geyer},
  {Giacobbe}, {Gilmore}, {Girona}, {Giuffrida}, {Gomel}, {Gomez},
  {Gonz{\'a}lez-N{\'u}{\~n}ez}, {Gonz{\'a}lez-Santamar{\'\i}a},
  {Gonz{\'a}lez-Vidal}, {Granvik}, {Guillout}, {Guiraud},
  {Guti{\'e}rrez-S{\'a}nchez}, {Guy}, {Hatzidimitriou}, {Hauser}, {Haywood},
  {Helmer}, {Helmi}, {Sarmiento}, {Hidalgo}, {Hilger}, {H{\l}adczuk}, {Hobbs},
  {Holland}, {Huckle}, {Jardine}, {Jasniewicz}, {Jean-Antoine Piccolo},
  {Jim{\'e}nez-Arranz}, {Jorissen}, {Juaristi Campillo}, {Julbe}, {Karbevska},
  {Kervella}, {Khanna}, {Kontizas}, {Kordopatis}, {Korn}, {K{\'o}sp{\'a}l},
  {Kostrzewa-Rutkowska}, {Kruszy{\'n}ska}, {Kun}, {Laizeau}, {Lambert},
  {Lanza}, {Lasne}, {Le Campion}, {Lebreton}, {Lebzelter}, {Leccia}, {Leclerc},
  {Lecoeur-Taibi}, {Liao}, {Licata}, {Lindstr{\o}m}, {Lister}, {Livanou},
  {Lobel}, {Lorca}, {Loup}, {Madrero Pardo}, {Magdaleno Romeo}, {Managau},
  {Mann}, {Manteiga}, {Marchant}, {Marconi}, {Marcos}, {Marcos Santos},
  {Mar{\'\i}n Pina}, {Marinoni}, {Marocco}, {Marshall}, {Polo},
  {Mart{\'\i}n-Fleitas}, {Marton}, {Mary}, {Masip}, {Massari},
  {Mastrobuono-Battisti}, {Mazeh}, {McMillan}, {Messina}, {Michalik}, {Millar},
  {Mints}, {Molina}, {Molinaro}, {Moln{\'a}r}, {Monari}, {Mongui{\'o}},
  {Montegriffo}, {Montero}, {Mor}, {Mora}, {Morbidelli}, {Morel}, {Morris},
  {Muraveva}, {Murphy}, {Musella}, {Nagy}, {Noval}, {Oca{\~n}a}, {Ogden},
  {Ordenovic}, {Osinde}, {Pagani}, {Pagano}, {Palaversa}, {Palicio},
  {Pallas-Quintela}, {Panahi}, {Payne-Wardenaar}, {Pe{\~n}alosa Esteller},
  {Penttil{\"a}}, {Pichon}, {Piersimoni}, {Pineau}, {Plachy}, {Plum}, {Poggio},
  {Pr{\v{s}}a}, {Pulone}, {Racero}, {Ragaini}, {Rainer}, {Raiteri}, {Rambaux},
  {Ramos}, {Ramos-Lerate}, {Re Fiorentin}, {Regibo}, {Richards}, {Rios Diaz},
  {Ripepi}, {Riva}, {Rix}, {Rixon}, {Robichon}, {Robin}, {Robin}, {Roelens},
  {Rogues}, {Rohrbasser}, {Romero-G{\'o}mez}, {Rowell}, {Royer}, {Ruz Mieres},
  {Rybicki}, {Sadowski}, {S{\'a}ez N{\'u}{\~n}ez}, {Sagrist{\`a} Sell{\'e}s},
  {Sahlmann}, {Salguero}, {Samaras}, {Sanchez Gimenez}, {Sanna},
  {Santove{\~n}a}, {Sarasso}, {Schultheis}, {Sciacca}, {Segol}, {Segovia},
  {S{\'e}gransan}, {Semeux}, {Shahaf}, {Siddiqui}, {Siebert}, {Siltala},
  {Silvelo}, {Slezak}, {Slezak}, {Smart}, {Snaith}, {Solano}, {Solitro},
  {Souami}, {Souchay}, {Spagna}, {Spina}, {Spoto}, {Steele},
  {Steidelm{\"u}ller}, {Stephenson}, {S{\"u}veges}, {Surdej}, {Szabados},
  {Szegedi-Elek}, {Taris}, {Taylo}, {Teixeira}, {Tolomei}, {Tonello}, {Torra},
  {Torra}, {Torralba Elipe}, {Trabucchi}, {Tsounis}, {Turon}, {Ulla}, {Unger},
  {Vaillant}, {van Dillen}, {van Reeven}, {Vanel}, {Vecchiato}, {Viala},
  {Vicente}, {Voutsinas}, {Weiler}, {Wevers}, {Wyrzykowski}, {Yoldas}, {Yvard},
  {Zhao}, {Zorec}, {Zucker}, \& {Zwitter}}]{Vallenari22}
{Vallenari}, A., {Brown}, A.~G.~A., {Prusti}, T., {et~al.} 2022, arXiv
  e-prints, arXiv:2208.00211

\bibitem[{{van den Eijnden} {et~al.}(2017){van den Eijnden}, {Bagnoli},
  {Degenaar}, {Lohfink}, {Parker}, {in 't Zand}, \& {Fabian}}]{vandenEijnden17}
{van den Eijnden}, J., {Bagnoli}, T., {Degenaar}, N., {et~al.} 2017, \mnras,
  466, L98

\bibitem[{{van der Klis}(2006)}]{vanderklis06}
{van der Klis}, M. 2006, in Compact stellar X-ray sources, Vol.~39, 39--112

\bibitem[{{van Paradijs}(1998)}]{vanParadijs98}
{van Paradijs}, J. 1998, in NATO Advanced Study Institute (ASI) Series C, Vol.
  515, The Many Faces of Neutron Stars., ed. R.~{Buccheri}, J.~{van Paradijs},
  \& A.~{Alpar}, 279

\bibitem[{{Wachter} {et~al.}(1979){Wachter}, {Leach}, \& {Kellogg}}]{Wachter79}
{Wachter}, K., {Leach}, R., \& {Kellogg}, E. 1979, \apj, 230, 274

\bibitem[{{Wijnands} \& {van der Klis}(2000)}]{Wijnands00}
{Wijnands}, R. \& {van der Klis}, M. 2000, \apjl, 528, L93

\bibitem[{{Wilms} {et~al.}(2000){Wilms}, {Allen}, \& {McCray}}]{Wilms00}
{Wilms}, J., {Allen}, A., \& {McCray}, R. 2000, \apj, 542, 914

\end{thebibliography}

\end{document}